\documentclass[onecolumn,11pt, notitlepage]{article}

\usepackage{amsmath,amsfonts,amssymb,amsthm}
\usepackage{graphicx}
\usepackage[labelsep=endash,labelfont=bf,justification=justified]{caption}
\usepackage[top=1in, bottom=1in, left=1in, right=1in]{geometry}
\usepackage[pdftex]{hyperref}

\allowdisplaybreaks

\usepackage[font=normal]{subcaption}

\usepackage[square,numbers]{natbib}
\begin{document}

\title{Model order reduction for stochastic \\ dynamical systems with continuous symmetries}
\date{\today}
\author{Saviz Mowlavi and Themistoklis P. Sapsis \\ \rule{0pt}{15pt} Department of Mechanical Engineering, Massachusetts Institute of \\ Technology, 77 Massachusetts Avenue, Cambridge MA 02139}

\maketitle

\begin{abstract}
Stochastic dynamical systems with continuous symmetries arise commonly in nature and often give rise to coherent spatio-temporal patterns. However, because of their random locations, these patterns are not well captured by current order reduction techniques and a large number of modes is typically necessary for an accurate solution. In this work, we introduce a new methodology for efficient order reduction of such systems by combining (i) the method of slices \cite{rowley2000,froehlich2012}, a symmetry reduction tool, with (ii) any standard order reduction technique, resulting in efficient mixed symmetry-dimensionality reduction schemes. In particular, using the Dynamically Orthogonal (DO) equations \cite{sapsis2009} in the second step, we obtain a novel nonlinear Symmetry-reduced Dynamically Orthogonal (SDO) scheme. We demonstrate the performance of the SDO scheme on stochastic solutions of the 1D Korteweg-de Vries and 2D Navier-Stokes equations.
\end{abstract}

\section{Introduction}

Examples of physical systems that can be modeled as stochastic dynamical systems with continuous symmetries abound in the world around us, whether it be pipe flow \cite{hof2004}, water waves \cite{yuen1980}, flame dynamics \cite{renard2000}, or nonlinear optics \cite{arecchi1999}, just to name a few. Stochasticity arises from unknown parameters or initial conditions, while continuous symmetry manifests itself as translational and/or rotational invariance due to specific geometry. Such systems are typically described by stochastic partial differential equations (PDEs) with complex nonlinear responses, which makes accurate quantification of their statistical behavior through direct Monte-Carlo simulations a challenge due to the high computational costs involved.

Model order reduction aims at solving this issue by approximating the stochastic solution in terms of a finite sum of deterministic spatial modes multiplied by stochastic scalar coefficients, motivated by the Karhunen-Lo\`eve decomposition. In this way, the computation of the stochastic solution is reduced to the evolution of the coefficients and/or the modes, leading to significant computational savings and, in some circumstances, to improved physical understanding of the underlying dynamics. Naturally, there red different possible ways to derive such reduced-order models from the governing equations, and various dimensionality reduction methods have been proposed over the years, such as the Proper Orthogonal Decomposition (POD) combined with Galerkin projection \cite{berkooz1993,holmes1996}, the Polynomial Chaos (PC) expansion \cite{xiu2002}, or more recently the Dynamically Orthogonal (DO) equations \cite{sapsis2009}.

Stochastic dynamical systems with continuous symmetries often give rise to coherent spatio-temporal patterns \cite{cross1993}. However, due to the fact that individual realizations are invariant along the symmetry directions of the system, the precise location of these spatio-temporal patterns might be subject to large stochastic variability and different realizations might display similar structures at completely different spatial locations. As a result, reduced-order models relying on linear modal decompositions will require a large number of spatial modes to adequately capture the spatio-temporal dynamics of the stochastic solution, which is likely to offset the computational gains. This is because spatial shifts cannot be efficiently represented by a finite linear combination of global spatial modes.

In parallel with dimensionality reduction methods, much work has been done over the last few decades on symmetry reduction, which concerns the removal of continuous symmetries associated with deterministic dynamical systems \cite{cartan1935,fels1998,rowley2000,beyn2004,siminos2011,froehlich2012,kreilos2014}. After symmetry reduction, the dynamics of the original system along its symmetry directions (i.e.~translations and/or rotations) is factored out, so that the resulting symmetry-reduced state is left with non-trivial shape-changing dynamics. Coherent spatio-temporal patterns appearing in the symmetry-reduced solution will therefore remain at the same spatial location while undergoing shape deformations. 

In this way, the possibly low-rank structure of the symmetry-reduced state is preserved, which is clearly advantageous for model order reduction purposes. \cite{kirby1992,glavaski1998} first combined such symmetry reduction techniques with the POD to derive low-dimensional models for deterministic systems governed by PDEs. Their model, however, only described the symmetry-reduced dynamics and no attempt was made to recover the original system state. Such a closure was accomplished shortly thereafter in the same context by \cite{rowley2000} with a so-called ``reconstruction equation" for the symmetry coordinate, resulting in the first dynamical order reduction framework to take advantage of the continuous symmetries of a system. The same procedure was later adopted by \cite{ohlberger2013} in the context of parametric order reduction using reduced-basis methods. More generally, combinations of nonlinear mappings with reduced-order models have been explored by several authors lately \cite{iollo2014,cagniart2016,cagniart2017,mojgani2017,nair2017}.

Similar ideas of reducing the symmetry before performing order reduction have also been applied to stochastic dynamical systems, but these recent studies focused on data reduction \cite{sonday2013} or inference \cite{ravela2015}. Here, we introduce a new framework for efficient dynamical model order reduction of stochastic dynamical systems with continuous symmetries by combining symmetry reduction with order reduction methods. As we will see, this approach naturally leads to novel nonlinear reduced-order models that efficiently take advantage of the continuous symmetries of the system, leading to much improved accuracy for a given number of modes. This methodology can be applied to any dimensionality reduction method of choice provided the latter preserves the symmetry reduction properties, which is true for techniques like the POD or the DO equations. We will illustrate our approach with the DO equations, which will result in a novel Symmetry-reduced Dynamically Orthogonal (SDO) scheme.

The paper is structured as follows. The general symmetry-dimensionality reduction methodology as well as the derivation of the SDO scheme are presented in Section \ref{sec:BlendingSymmetryReductionWithDimensionalityReduction}. The performance of the SDO scheme is compared with the standard DO method on stochastic simulations of the 1D Korteweg-de Vries and 2D Navier-Stokes equations in Sections \ref{sec:ApplicationKdVEquation} and \ref{sec:ApplicationNavierStokesEquations}, respectively. Finally, conclusions follow in Section \ref{sec:Conclusions}.

\section{Blending symmetry reduction with dimensionality reduction}
\label{sec:BlendingSymmetryReductionWithDimensionalityReduction}

Let $(\Omega,\mathcal{B},\mathcal{P})$ be a probability space and $\omega \in \Omega$ indicate an elementary event. Denoting space $\mathbf{x} \in D \subset \mathbb{R}^n$ and time $t$, we consider the stochastic partial differential equation
\begin{equation}
\frac{\partial \mathbf{u}}{\partial t} = \mathbf{F}(\mathbf{u},t;\omega),
\label{eq:GoverningEquation}
\end{equation}
where $\mathbf{F}$ is a (possibly stochastic and time-dependent) nonlinear differential operator and $\mathbf{u}(\mathbf{x},t;\omega)$ is a random vector field that belongs to the Hilbert space $\mathcal{H}$ of continuous and square-integrable functions with inner product
\begin{equation}
\langle \mathbf{u}_1, \mathbf{u}_2 \rangle = \int_D \mathbf{u}_1 \mathbf{u}_2^* \, \mathrm{d}\mathbf{x}.
\label{eq:InnerProduct}
\end{equation}
In this work, we are interested in dynamical systems that are symmetric under a group $G$ of continuous transformations, that is, the differential operator $\mathbf{F}$ satisfies the following equivariance condition for any group element $g \in G$
\begin{equation}
\mathbf{F}(g \mathbf{u}) = g \mathbf{F}(\mathbf{u}).
\label{eq:EquivarianceCondition}
\end{equation}
We will restrict ourselves to symmetry groups $G$ that (i) are Lie groups, (ii) preserve inner products and distances. Typically, $G$ will consist of translations along different direction and/or rotations about different axes. When the above relation \eqref{eq:EquivarianceCondition} is satisfied, individual solutions $\mathbf{u}$ of the dynamical system \eqref{eq:GoverningEquation} are invariant under $G$, meaning that $g \mathbf{u}$ is also a solution for any $g \in G$.

In this paper, we introduce a new framework for efficient dimensionality reduction of such systems with continuous symmetries. Our methodology comprises the two following steps:
\begin{enumerate}
\item In a first step, we perform \textbf{symmetry reduction} of the dynamical system \eqref{eq:GoverningEquation} using the \textit{method of slices} \cite{rowley2000,froehlich2012}. In this framework, the original system state $\mathbf{u}$ is decomposed into (i) a stochastic symmetry-reduced state $\hat{\mathbf{u}}$, which is fixed in the physical domain but captures the intrinsic changes in shape of $\mathbf{u}$, and (ii) a finite set of stochastic phase parameters $\boldsymbol{\phi}$ that track the motion of $\mathbf{u}$ along the symmetry directions of the system. The symmetry-reduced state $\hat{\mathbf{u}}$ and phase parameters $\boldsymbol{\phi}$ are defined precisely in Section \ref{sec:MethodSlices}, and equations governing their temporal evolution are presented in Section \ref{sec:DynamicsWithinSymmetryReducedStateSpace}.
\item In a second step, we perform \textbf{dimensionality reduction} of the stochastic symmetry-reduced state using any standard model order reduction method of choice. This order reduction step is illustrated in Section \ref{sec:OrderReductionSymmetryReducedState} through the use of the Dynamically Orthogonal (DO) equations \cite{sapsis2009}, which leads to a novel, nonlinear Symmetry-reduced Dynamically Orthogonal (SDO) scheme.
\end{enumerate}

In the following, we adopt a dynamical state-space approach where the system state $\mathbf{u}(t;\omega)$ for given time $t$ and realization $\omega$ is represented by a single point in the infinite dimensional state space $\mathcal{M}$ of all possible solutions.

\subsection{Method of slices}
\label{sec:MethodSlices}

\begin{figure}
\centering
\includegraphics[scale=0.8]{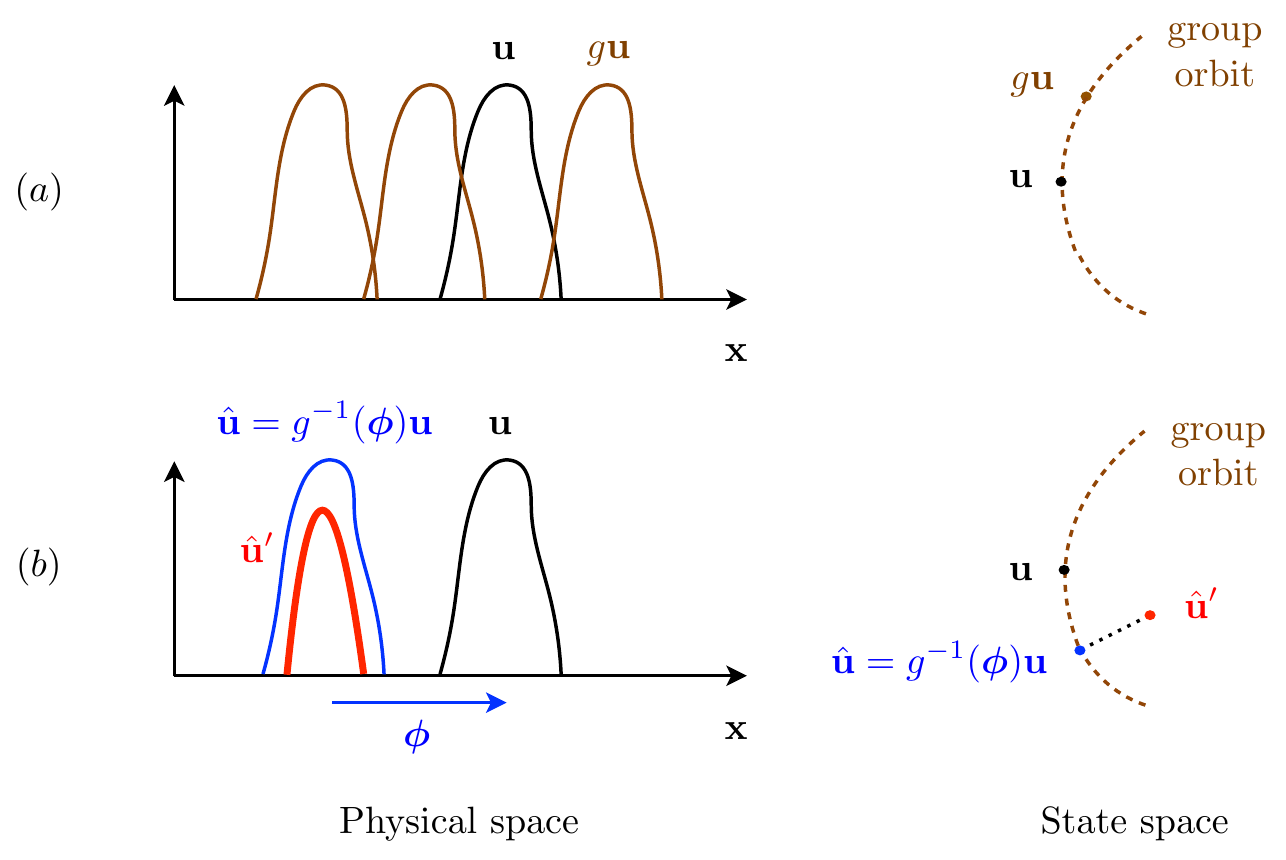}
\caption{Symmetry reduction with the method of slices applied to a given state $\mathbf{u}$, for instance a particular realization $\omega_0$ at a given time $t_0$ of a dynamical system with translational symmetry. (a) Applying the family of continuous transformations $g \in G$ to $\mathbf{u}$ gives rise to a family of dynamically equivalent states $g\mathbf{u}$ called the \textit{group orbit} of $\mathbf{u}$. (b) The method of slices defines $\hat{\mathbf{u}}$ by choosing the point on the group orbit of $\mathbf{u}$ that minimizes the distance $|| \hat{\mathbf{u}} - \hat{\mathbf{u}}'||$ to a fixed template state $\hat{\mathbf{u}}'$. Writing $\hat{\mathbf{u}} = g^{-1}(\boldsymbol{\phi}) \mathbf{u}$, this minimum distance condition is achieved when the phase parameters $\boldsymbol{\phi}$ satisfy the slice condition \eqref{eq:SliceCondition2}.}
\label{fig:MethodSlices}
\end{figure}

\begin{figure}
\centering
\includegraphics[scale=0.8]{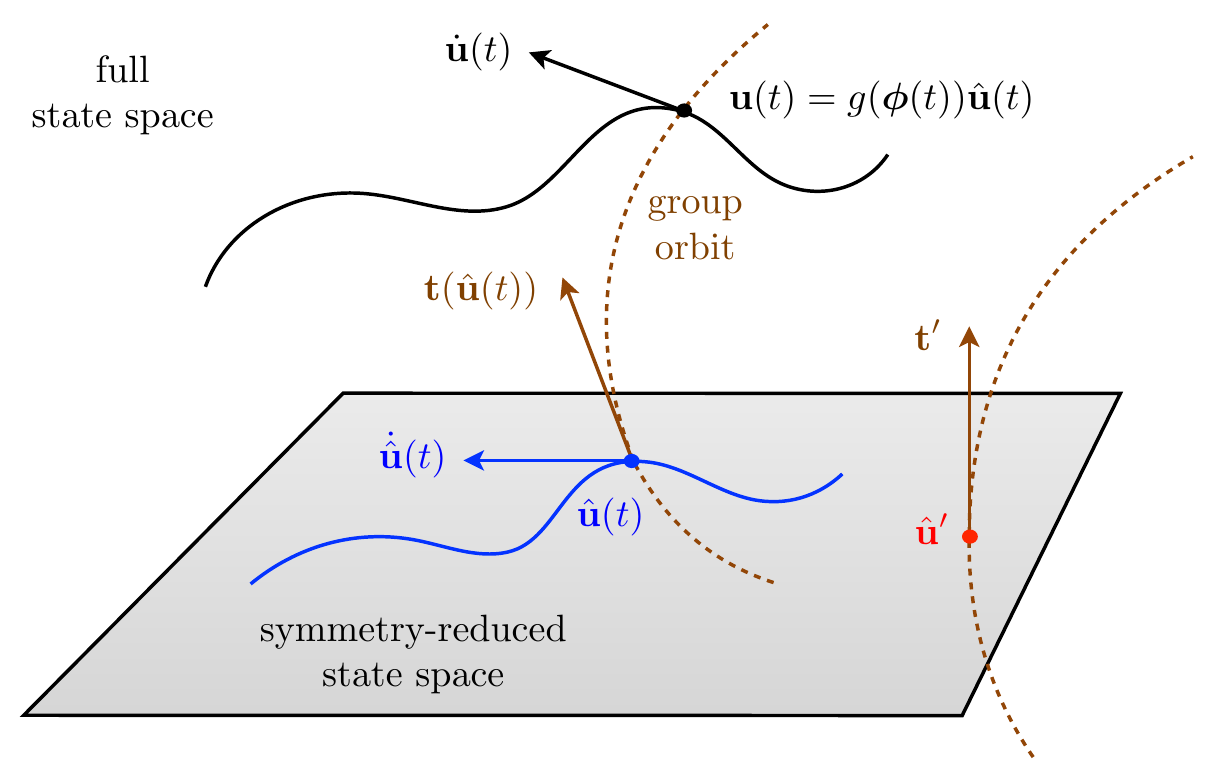}
\caption{Symmetry reduction with the method of slices applied to a stochastic dynamical system, shown here for the time trajectory of a particular realization $\mathbf{u}(t) = \mathbf{u}(t;\omega_0)$ with corresponding symmetry-reduced state $\hat{\mathbf{u}}(t) = \hat{\mathbf{u}}(t;\omega_0)$ and phase parameters $\boldsymbol{\phi}(t)$ = $\boldsymbol{\phi}(t;\omega_0)$. While $\mathbf{u}(t)$ is evolving in the full state space under the governing equation \eqref{eq:GoverningEquation}, $\hat{\mathbf{u}}(t)$ and $\boldsymbol{\phi}(t)$ evolve under equations \eqref{eq:ReducedStateDynamics} and \eqref{eq:PhaseParametersDynamics} in such a way that the slice condition \eqref{eq:SliceCondition2} is satisfied at all times $t$. As a result, the trajectory of $\hat{\mathbf{u}}(t)$ remains confined to the symmetry-reduced state space passing through the fixed template $\hat{\mathbf{u}}'$ and orthogonal to its group orbit tangents $\mathbf{t}_a' = \mathbf{t}_a(\hat{\mathbf{u}}')$. One can always reconstruct the full stochastic state from the symmetry-reduced stochastic dynamics as $\mathbf{u}(t;\omega) = g(\boldsymbol{\phi}(t;\omega)) \hat{\mathbf{u}}(t;\omega)$.}
\label{fig:MethodSlicesDynamics}
\end{figure}

Let us first consider a given deterministic state $\mathbf{u}$, for instance a particular realization $\omega_0$ at a given time $t_0$ of \eqref{eq:GoverningEquation}. The application of the family of continuous transformations $g \in G$ to $\mathbf{u}$ gives rise to a family of dynamically equivalent states $g\mathbf{u}$ called the \textit{group orbit} of $\mathbf{u}$, as illustrated in Figure \ref{fig:MethodSlices}(a).
The goal of symmetry reduction is to reduce all these equivalent symmetry copies to a unique representative symmetry-reduced state $\hat{\mathbf{u}}$. For a given state $\mathbf{u}$ and its group orbit, the \textit{method of slices} defines $\hat{\mathbf{u}}$ by choosing the point on the group orbit of $\mathbf{u}$ that is closest to a fixed \textit{template} state $\hat{\mathbf{u}}'$, so that $\hat{\mathbf{u}}$ overlies as well as possible the template in physical space, as indicated in Figure \ref{fig:MethodSlices}(b). Denoting with $\boldsymbol{\phi} = \phi_1, ..., \phi_N$ the $N$ continuous scalar phase parameters of the transformation $G$ (for example the translations amounts and/or rotation angles along different directions) and writing $\mathbf{u} = g(\boldsymbol{\phi}) \hat{\mathbf{u}}$ or equivalently $\hat{\mathbf{u}} = g^{-1}(\boldsymbol{\phi}) \mathbf{u}$, this condition can be expressed in terms of $\boldsymbol{\phi}$ as
\begin{equation}
\min_{\boldsymbol{\phi}} || g^{-1}(\boldsymbol{\phi}) \mathbf{u} - \hat{\mathbf{u}}' ||,
\label{eq:MinimizationProblem}
\end{equation}
where we use the $L^2$ norm $||\mathbf{u}||^2 = \langle \mathbf{u},\mathbf{u} \rangle$. As mentioned earlier, we only consider transformations that do not affect the inner product between different states, i.e.~$\langle g\mathbf{u}_1, g\mathbf{u}_2 \rangle = \langle \mathbf{u}_1, \mathbf{u}_2 \rangle$ and $|| g\mathbf{u} || = || \mathbf{u} ||$. Therefore, the above condition becomes
\begin{equation}
\min_{\boldsymbol{\phi}} || \mathbf{u} - g(\boldsymbol{\phi}) \hat{\mathbf{u}}' ||,
\end{equation}
from which we can deduce the extremum condition
\begin{equation}
\frac{\partial}{\partial \phi_a} || \mathbf{u} - g(\boldsymbol{\phi}) \hat{\mathbf{u}}' ||^2 = \frac{\partial}{\partial \phi_a} \langle \mathbf{u}, \mathbf{u} \rangle - 2 \frac{\partial}{\partial \phi_a} \langle \mathbf{u}, g(\boldsymbol{\phi})\hat{\mathbf{u}}' \rangle + \frac{\partial}{\partial \phi_a} \langle g(\boldsymbol{\phi})\hat{\mathbf{u}}', g(\boldsymbol{\phi})\hat{\mathbf{u}}' \rangle = 0,
\end{equation}
where $a = 1,...,N$. Strictly speaking, one should also make sure that the second derivative with respect to $\phi_a$ be positive in order to have a minimum. Using the distance-preserving property of the transformation, the extremum condition leads to the following \textit{slice condition}
\begin{equation}
\langle \mathbf{u}, \mathbf{t}_a(g(\boldsymbol{\phi})\hat{\mathbf{u}}') \rangle = 0,
\label{eq:SliceCondition}
\end{equation}
where $\mathbf{t}_a(g(\boldsymbol{\phi})\hat{\mathbf{u}}')$ is the tangent to the group orbit at $g(\boldsymbol{\phi})\hat{\mathbf{u}}'$ in direction $\phi_a$
\begin{equation}
\mathbf{t}_a(g(\boldsymbol{\phi})\hat{\mathbf{u}}') = \frac{\partial g(\boldsymbol{\phi})\hat{\mathbf{u}}'}{\partial \phi_a} = \lim_{\delta\phi_a \rightarrow 0} \frac{g(\boldsymbol{\phi}+\delta\phi_a)\hat{\mathbf{u}}' - g(\boldsymbol{\phi})\hat{\mathbf{u}}'}{\delta\phi_a}.
\label{eq:Generator}
\end{equation}
Since $G$ is a Lie group, we have $g(\boldsymbol{\phi}+\delta\phi_a)\hat{\mathbf{u}}' = g(\boldsymbol{\phi}) g(\delta\phi_a)\hat{\mathbf{u}}'$, thus we can factor out the group action $g(\boldsymbol{\phi})$ from the above expression to obtain the relation
\begin{equation}
\mathbf{t}_a(g(\boldsymbol{\phi})\hat{\mathbf{u}}') = g(\boldsymbol{\phi}) \mathbf{t}_a(\hat{\mathbf{u}}'),
\end{equation}
where $\mathbf{t}_a(\hat{\mathbf{u}}')$ is the group orbit tangent at the fixed template $\hat{\mathbf{u}}'$ in direction $\phi_a$
\begin{equation}
\mathbf{t}_a(\hat{\mathbf{u}}') = \lim_{\delta\phi_a \rightarrow 0} \frac{g(\delta\phi_a)\hat{\mathbf{u}}' - \hat{\mathbf{u}}'}{\delta\phi_a}.
\end{equation}
As a result, we may now  use the distance-preserving property of $g$ to rewrite the slice condition \eqref{eq:SliceCondition} in terms of the symmetry-reduced state $\hat{\mathbf{u}}$
\begin{equation}
\langle \mathbf{u}, g(\boldsymbol{\phi})\mathbf{t}_a' \rangle = 0 \quad \Leftrightarrow \quad \langle \hat{\mathbf{u}}, \mathbf{t}_a' \rangle = 0,
\label{eq:SliceCondition2}
\end{equation}
where we have denoted the fixed template tangent $\mathbf{t}_a' = \mathbf{t}_a(\hat{\mathbf{u}}')$. Given a state $\mathbf{u}$ and template $\hat{\mathbf{u}}'$, the first equality in the above slice condition gives the phase parameters $\boldsymbol{\phi}$ such that the distance between the symmetry-reduced state $\hat{\mathbf{u}} = g^{-1}(\boldsymbol{\phi}) \mathbf{u}$ and the template $\hat{\mathbf{u}}'$ is minimized. The second equality can be interpreted as an orthogonality condition which states that the symmetry-reduced state $\hat{\mathbf{u}}$ always lies within a hyperplane normal to the $N$ group orbit tangents $\mathbf{t}_a'$ at the template $\hat{\mathbf{u}}'$. This fixed hyperplane thus defines a slice through the full state space containing all symmetry-reduced states and called the \textit{symmetry-reduced state space}, see Figure \ref{fig:MethodSlicesDynamics}.

\subsection{Dynamics within the symmetry-reduced state space}
\label{sec:DynamicsWithinSymmetryReducedStateSpace}

Consider now that $\mathbf{u}(t;\omega)$ is stochastic and evolves under the governing equation \eqref{eq:GoverningEquation}, as illustrated in Figure \ref{fig:MethodSlicesDynamics}. 
The evolution of $\mathbf{u}(t;\omega)$ in the full state space will correspond a set of stochastic and time-dependent phase parameters $\boldsymbol{\phi}(t;\omega)$ and a stochastic symmetry-reduced state $\hat{\mathbf{u}}(t;\omega) = g^{-1}(\boldsymbol{\phi}(t;\omega)) \mathbf{u}(t;\omega)$ evolving in the symmetry-reduced state space in such a way that the slice condition \eqref{eq:SliceCondition2} is satisfied at all times and for all realizations. 
In other words, the symmetry-reduced state is fixed to the template in physical space and captures the shape deformations of $\mathbf{u}(t;\omega)$, while the phase parameters track the motion of $\mathbf{u}(t;\omega)$ along the symmetry directions of the system. In order to find the evolution equations for the symmetry-reduced state $\hat{\mathbf{u}}$ and the phase parameters $\boldsymbol{\phi}$, we first plug the relation $\mathbf{u} = g(\boldsymbol{\phi}) \hat{\mathbf{u}}$ into the governing equations
\begin{equation}
\dot{\phi}_a \frac{\partial g(\boldsymbol{\phi})\hat{\mathbf{u}}}{\partial \phi_a} + g(\boldsymbol{\phi}) \frac{\partial \hat{\mathbf{u}}}{\partial t} = \mathbf{F}(g(\boldsymbol{\phi}) \hat{\mathbf{u}}),
\end{equation}
where repeated indices indicate summation. As was done in \eqref{eq:Generator}, we next express the term $\partial g(\boldsymbol{\phi})\hat{\mathbf{u}}/ \partial \phi_a$ as
\begin{equation}
\frac{\partial g(\boldsymbol{\phi})\hat{\mathbf{u}}}{\partial \phi_a} = \mathbf{t}_a(g(\boldsymbol{\phi})\hat{\mathbf{u}}) = g(\boldsymbol{\phi}) \mathbf{t}_a(\hat{\mathbf{u}}),
\end{equation}
where $\mathbf{t}_a(\hat{\mathbf{u}})$ is the group orbit tangent of the time-dependent state $\hat{\mathbf{u}}$ in direction $\phi_a$
\begin{equation}
\mathbf{t}_a(\hat{\mathbf{u}}) = \lim_{\delta\phi_a \rightarrow 0} \frac{g(\delta\phi_a)\hat{\mathbf{u}} - \hat{\mathbf{u}}}{\delta\phi_a}.
\end{equation}
Finally, we make use of the equivariance condition \eqref{eq:EquivarianceCondition} to obtain the following evolution equation for the symmetry-reduced state
\begin{equation}
\frac{\partial \hat{\mathbf{u}}}{\partial t} = \mathbf{F}(\hat{\mathbf{u}}) - \dot{\phi}_a \mathbf{t}_a(\hat{\mathbf{u}}).
\label{eq:ReducedStateDynamics}
\end{equation}
It remains to find evolution equations for the phase parameters $\boldsymbol{\phi}$, which is achieved by substituting the above equation into the time derivative of the slice condition \eqref{eq:SliceCondition2}
\begin{equation}
\langle \frac{\partial \hat{\mathbf{u}}}{\partial t}, \mathbf{t}_a' \rangle = \langle \mathbf{F}(\hat{\mathbf{u}}), \mathbf{t}_a' \rangle  - \dot{\phi}_b \, \langle \mathbf{t}_b(\hat{\mathbf{u}}), \mathbf{t}_a' \rangle = 0.
\label{eq:DynamicSliceCondition}
\end{equation}
The above equation can be interpreted as an orthogonality condition that constrains the dynamics of $\hat{\mathbf{u}}$ to remain confined within the symmetry-reduced state space defined by the fixed template $\hat{\mathbf{u}}'$. Introducing the stochastic time-dependent matrix $\{\mathbf{T}\}_{ab} = \langle \mathbf{t}_b(\hat{\mathbf{u}}), \mathbf{t}_a' \rangle$ and vector $\{\mathbf{f}\}_a = \langle \mathbf{F}(\hat{\mathbf{u}}), \mathbf{t}_a' \rangle$, one finally gets the following matrix inverse problem for the phase velocity
\begin{equation}
\dot{\boldsymbol{\phi}} = \mathbf{T}^{-1} \mathbf{f}.
\label{eq:PhaseParametersDynamics}
\end{equation}
Together, equations \eqref{eq:ReducedStateDynamics} and \eqref{eq:PhaseParametersDynamics} govern the evolution of the stochastic symmetry-reduced state $\hat{\mathbf{u}}(t;\omega)$ and the phase parameters $\boldsymbol{\phi}(t;\omega)$ (see also \cite{cvitanovic2016} in the deterministic context). In this way, the dynamics of $\mathbf{u}(t;\omega)$ has been separated into shape deformations, reproduced by $\hat{\mathbf{u}}(t;\omega)$ which is fixed in physical space, and motion along the symmetry directions of the system, tracked by $\boldsymbol{\phi}(t;\omega)$. From these two quantities, the evolution of the system in the full state space can be reconstructed exactly as $\mathbf{u}(t;\omega) = g(\boldsymbol{\phi}(t;\omega)) \hat{\mathbf{u}}(t;\omega)$, hence no information has been lost so far.

Before moving on to the next step, we note that the choice of template can affect the symmetry reduction procedure in two distinct ways. For a given state $\mathbf{u}$, different templates will naturally lead to different optimal values of $\boldsymbol{\phi}$ for which the symmetry-reduced state $\hat{\mathbf{u}} = g^{-1}(\boldsymbol{\phi}) \mathbf{u}$ is closest to the template $\hat{\mathbf{u}}'$. Nevertheless, the full state reconstructed from the solution to equations \eqref{eq:ReducedStateDynamics} and \eqref{eq:PhaseParametersDynamics} will be independent of the template.
The second and potentially more worrisome effect concerns the conditioning of the matrix $\mathbf{T}$, or worse, the possibility of having singularities in equation \eqref{eq:PhaseParametersDynamics} whenever the determinant of $\mathbf{T}$ vanishes. This is a consequence of the fact that there exist states $\mathbf{u}$ for which a solution to the minimization problem \eqref{eq:MinimizationProblem}, or equivalently the slice condition \eqref{eq:SliceCondition2}, ceases to exist (a trivial exemple being the spatially constant state). The number and nature of such problematic states depend on the template function and reflect the finite extent of validity of the symmetry-reduced state space defined by a given template.
Therefore, the well-posedness of \eqref{eq:PhaseParametersDynamics} and the likelihood of encountering singularities are critically tied to the choice of template. We will see in Sections \ref{sec:ChoiceTemplateKdV} and \ref{sec:ChoiceTemplateNS} that there exist clever choices which almost entirely alleviate this issue.

\subsection{Order reduction in the symmetry-reduced state space}
\label{sec:OrderReductionSymmetryReducedState}

\begin{figure}[tb]
\centering
\includegraphics[scale=0.8]{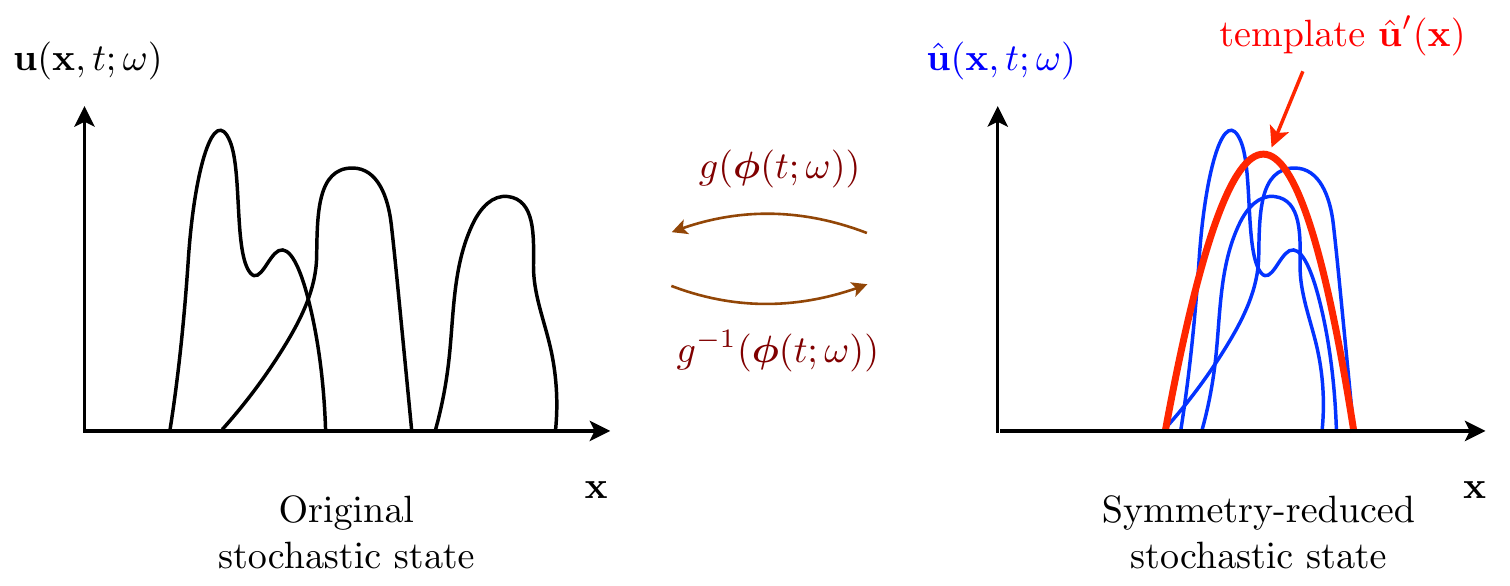}
\caption{A physical space representation of the method of slices applied to a stochastic system at a given time instant. Each realization $\mathbf{u}(\mathbf{x},t;\omega)$ of the original stochastic state is reduced to the symmetry-reduced realization $\hat{\mathbf{u}}(\mathbf{x},t;\omega) = g^{-1}(\boldsymbol{\phi}(t;\omega)) \mathbf{u}(\mathbf{x},t;\omega)$ that minimizes the distance $||\hat{\mathbf{u}}(\mathbf{x},t;\omega) - \hat{\mathbf{u}}'(\mathbf{x})||$, where $\hat{\mathbf{u}}'(\mathbf{x})$ is a fixed template. As a result, the symmetry-reduced realizations $\hat{\mathbf{u}}(\mathbf{x},t;\omega)$ can be approximated much better by low-dimensional models that rely on linear modal decompositions than their full state space counterparts $\mathbf{u}(\mathbf{x},t;\omega)$.}
\label{fig:SymmetryReduction}
\end{figure}

We now turn to the second step of our methodology, which consists in applying standard order reduction methods directly to the dynamics of $\hat{\mathbf{u}}$ in the symmetry-reduced space, governed by equations \eqref{eq:ReducedStateDynamics} and \eqref{eq:PhaseParametersDynamics}. Because the symmetry-reduced stochastic state is defined in such a way that the distance $||\hat{\mathbf{u}}(\mathbf{x},t;\omega) - \hat{\mathbf{u}}'(\mathbf{x})||$ is minimized at all times and for all realizations, the symmetry-reduced realizations $\hat{\mathbf{u}}(\mathbf{x},t;\omega)$ will be grouped together in physical space, as illustrated in Figure \ref{fig:SymmetryReduction}.
This property is very appealing for low-dimensional models that rely on linear modal decompositions since it implies that the symmetry-reduced realizations $\hat{\mathbf{u}}(\mathbf{x},t;\omega)$ can be approximated much better by such finite-dimensional modal decompositions than their full state space counterparts $\mathbf{u}(\mathbf{x},t;\omega)$. This is inherently connected to the notion of Kolmogorov \textit{n}-width from approximation theory \cite{kolmogorov1936,pinkus1985}, and we may say that the symmetry-reduced realizations have much lower \textit{n}-width than their full state space counterparts \cite{ohlberger2013,cagniart2016}.
 
 Here, as an illustration, we perform dimensionality reduction using the DO framework \cite{sapsis2009} because of its inherent and desirable ability to deal with strongly transient stochastic responses. Moreover, by considering the DO equations as a computational method for evolving a low-rank matrix representation of the discretized solution, \cite{feppon2017} recently showed that DO gives the best possible instantaneous approximation amongst reduced-order models relying on a linear modal decomposition. In general, we note that when choosing an order reduction method, one needs to make sure that the order reduction step preserves the symmetry reduction step, that is, the dynamics of the low-dimensional solution remain confined to the symmetry-reduced state space.

As discussed above, we apply the DO equations directly to the symmetry-reduced state $\hat{\mathbf{u}}$ instead of the original state $\mathbf{u}$, resulting in a new order reduction framework that we call the Symmetry-reduced Dynamically Orthogonal (SDO) equations. Following the standard DO methodology, the stochastic symmetry-reduced state is decomposed into a mean component, and a stochastic part that is projected to a low-dimensional subspace of order $s$ through the following finite-dimensional expansion
\begin{equation}
\hat{\mathbf{u}}(\mathbf{x},t;\omega) = \bar{\mathbf{u}}(\mathbf{x},t) + Y_i(t;\omega) \hat{\mathbf{u}}_i(\mathbf{x},t), \quad i = 1,...,s,
\label{eq:DORepresentation}
\end{equation}
where $\bar{\mathbf{u}}(\mathbf{x},t)$ is the mean of the symmetry-reduced state, $\hat{\mathbf{u}}_i(\mathbf{x},t)$ are time-dependent orthonormal modes that capture the principal directions of variance of the symmetry-reduced state, and $Y_i(t;\omega)$ are time-dependent stochastic coefficients that characterize the symmetry-reduced stochastic state within this subspace. The redundancy arising from the dependence in time of both the modes and stochastic coefficients is overcome through the DO condition
\begin{equation}
\langle \frac{\partial \hat{\mathbf{u}}_i}{\partial t}, \hat{\mathbf{u}}_j \rangle = 0, \quad i,j = 1,...,s,
\label{eq:DOCondition}
\end{equation}
which requires that the time variation of the stochastic subspace be orthogonal to itself. Inserting the DO representation \eqref{eq:DORepresentation} into equation \eqref{eq:ReducedStateDynamics} governing the evolution of $\hat{\mathbf{u}}(\mathbf{x},t;\omega)$ in the symmetry-reduced state space and using the DO condition \eqref{eq:DOCondition}, one obtains, together with equation \eqref{eq:PhaseParametersDynamics} for the phase parameters $\boldsymbol{\phi}(t;\omega)$, an explicit set of evolution equations for all unknown quantities. We have the following deterministic PDE for the mean field
\begin{equation}
\frac{\partial \bar{\mathbf{u}}}{\partial t} = E[\mathbf{F}(\hat{\mathbf{u}}) - \dot{\phi}_a \mathbf{t}_a(\hat{\mathbf{u}})],
\label{eq:DOMean}
\end{equation}
the following stochastic ODEs for the stochastic coefficients
\begin{equation}
\frac{\mathrm{d} Y_i}{\mathrm{d} t} = \langle \mathbf{F}(\hat{\mathbf{u}}) - \dot{\phi}_a \mathbf{t}_a(\hat{\mathbf{u}}) - E[\mathbf{F}(\hat{\mathbf{u}}) - \dot{\phi}_a \mathbf{t}_a(\hat{\mathbf{u}})], \hat{\mathbf{u}}_i \rangle,
\label{eq:DOCoefficients}
\end{equation}
and the following deterministic PDEs for the modes
\begin{equation}
\frac{\partial \hat{\mathbf{u}}_i}{\partial t} = \hat{\mathbf{H}}_i - \langle \hat{\mathbf{H}}_i, \hat{\mathbf{u}}_j \rangle \hat{\mathbf{u}}_j,
\label{eq:DOModes}
\end{equation}
where the deterministic fields $\hat{\mathbf{H}}_i$ are defined as
\begin{equation}
\hat{\mathbf{H}}_i = E[Y_k(\mathbf{F}(\hat{\mathbf{u}}) - \dot{\phi}_a \mathbf{t}_a(\hat{\mathbf{u}}))] \, C_{ik}^{-1},
\label{eq:DOHModes}
\end{equation}
where $C_{ij} = E[Y_i Y_j]$ is the covariance matrix. The continuous phase parameters $\boldsymbol{\phi}(t;\omega)$ are evolved according to the stochastic equation \eqref{eq:PhaseParametersDynamics}
\begin{equation}
\dot{\boldsymbol{\phi}} = \mathbf{T}^{-1} \mathbf{f},
\label{eq:DOPhaseParameters}
\end{equation}
where $\{\mathbf{T}\}_{ab} = \langle \mathbf{t}_b(\hat{\mathbf{u}}), \mathbf{t}_a' \rangle$ and $\{\mathbf{f}\}_a = \langle \mathbf{F}(\hat{\mathbf{u}}), \mathbf{t}_a' \rangle$. Altogether, equations \eqref{eq:DOMean}, \eqref{eq:DOCoefficients}, \eqref{eq:DOModes} and \eqref{eq:DOPhaseParameters} constitute the SDO scheme. They entirely specify the time evolution of the symmetry-reduced state $\hat{\mathbf{u}}(\mathbf{x},t;\omega)$ and the phase parameters $\boldsymbol{\phi}(t;\omega)$. In Appendix \ref{app:OrderReductionPreservesSymmetryReduction}, we show that the dynamics of the reduced-order solution \eqref{eq:DORepresentation} remain confined to the symmetry-reduced state space. Finally, the full state is easily recovered through the group transformation
\begin{equation}
\mathbf{u}(\mathbf{x},t;\omega) = g(\boldsymbol{\phi}(t;\omega)) \hat{\mathbf{u}}(\mathbf{x},t;\omega) = g(\boldsymbol{\phi}(t;\omega))\bar{\mathbf{u}}(\mathbf{x},t) + Y_i(t;\omega) g(\boldsymbol{\phi}(t;\omega))\hat{\mathbf{u}}_i(\mathbf{x},t).
\end{equation}
The above equation shows that the SDO framework can be interpreted as a \textit{nonlinear} extension of the usual DO methodology, wherein the mean and modes are allowed further degrees of freedom through the stochastic and time-dependent group transformation $g(\boldsymbol{\phi}(t;\omega))$ along the symmetry directions $G$ of the system. In this sense, it is naturally expected that SDO performs as well or better than DO, which we will see in the following sections where we apply both schemes to concrete one and two-dimensional examples.

\section{Application to the KdV equation}
\label{sec:ApplicationKdVEquation}

We first illustrate our approach with the Korteweg-de Vries (KdV) equation that describes the evolution of weakly nonlinear waves on shallow water surfaces
\begin{equation}
\frac{\partial u}{\partial t} + u \frac{\partial u}{\partial x} + \mu \frac{\partial^3 u}{\partial x^3} = 0,
\label{eq:KdV}
\end{equation}
where $u(x,t)$ is the surface elevation, $x \in [0,L]$ the space variable and $t$ the time. Stochasticity will be introduced through the initial conditions. The boundary conditions are periodic so that the KdV equation is equivariant under the symmetry group of continuous translations, $G = SO(2)_x$. The associated shift operator $g \in G$ writes
\begin{equation}
g(c) u(x) = u(x-c),
\end{equation}
where the continuous phase parameter $c$ represents the shift amount. The tangent to the group orbit at an arbitrary state $u$ is then given by
\begin{equation}
t(u) = \lim_{\delta c \rightarrow 0} \frac{g(\delta c)u - u}{\delta c} = -\frac{\partial u}{\partial x}.
\end{equation}

\subsection{Dynamics within the symmetry-reduced state space} 

As described in Section \ref{sec:DynamicsWithinSymmetryReducedStateSpace}, we now pick a fixed template $\hat{u}'(x)$ and we consider the evolution of the symmetry-reduced stochastic state $\hat{u}(x,t;\omega) = g^{-1}(c(t;\omega)) u(x,t;\omega)$ such that the distance $||\hat{u}(x,t;\omega) - \hat{u}'(x)||$ between the reduced state and the template is minimized at all times $t$ and for all realizations $\omega$. In the case of the KdV equation subject to the symmetry group $G$, the dynamical equations \eqref{eq:ReducedStateDynamics} for $\hat{u}$ in the symmetry-reduced state space become
\begin{equation}
\frac{\partial \hat{u}}{\partial t} = F(\hat{u}) + \dot{c} \, \frac{\partial \hat{u}}{\partial x},
\label{eq:ReducedStateDynamicsKdV}
\end{equation}
where $F(u)$ is the differential operator corresponding to the KdV equation in the full state space
\begin{equation}
F(u) = - u \frac{\partial u}{\partial x} - \mu \frac{\partial^3 u}{\partial x^3},
\end{equation}
and the stochastic differential equation \eqref{eq:PhaseParametersDynamics} for the shift amount $c(t;\omega)$ writes 
\begin{equation}
\dot{c} = \frac{\langle F(\hat{u}), t' \rangle}{\langle t(\hat{u}), t' \rangle},
\label{eq:PhaseParametersDynamicsKdV}
\end{equation}
where $t' = t(\hat{u}')$ is the group orbit tangent to the fixed template $\hat{u}'$.

\subsection{Order reduction in the symmetry-reduced state space}

In order to formulate the reduced-order SDO equations, let us first approximate the symmetry-reduced stochastic state as a truncated Karhunen-Loeve expansion
\begin{equation}
\hat{u}(x,t;\omega) = \bar{u}(x,t) + Y_i(t;\omega) \hat{u}_i(x,t), \quad i = 1,...,s,
\label{eq:DORepresentationKdV}
\end{equation}
where $\bar{u}$ and $\hat{u}_i$ are respectively the symmetry-reduced mean and modes and $Y_i$ are the stochastic coefficients of the finite-dimensional expansion. Next, we insert the above representation into the RHS of the evolution equation \eqref{eq:ReducedStateDynamicsKdV} for the symmetry-reduced state
\begin{equation}
F(\hat{u}) + \dot{c} \, \frac{\partial \hat{u}}{\partial x} = F_0 + Y_i \, F_i + Y_i Y_j \, F_{ij} + \dot{c} \, \frac{\partial \bar{u}}{\partial x} + \dot{c} \, Y_i \frac{\partial \hat{u}_i}{\partial x},
\label{eq:ExpandedOperatorKdV}
\end{equation}
where $F_0$, $F_i$, and $F_{ij}$ are deterministic fields given by
\begin{subequations}
\begin{gather}
F_0 = - \bar{u} \frac{\partial \bar{u}}{\partial x} - \mu \frac{\partial^3 \bar{u}}{\partial x^3}, \\
F_i = - \bar{u} \frac{\partial \hat{u}_i}{\partial x} - \hat{u}_i \frac{\partial \bar{u}}{\partial x} - \mu \frac{\partial^3 \hat{u}_i}{\partial x^3}, \\
F_{ij} = - \hat{u}_i \frac{\partial \hat{u}_j}{\partial x}.
\end{gather}
\end{subequations}
Now, we substitute the expanded RHS operator \eqref{eq:ExpandedOperatorKdV} into the SDO equations \eqref{eq:DOMean}, \eqref{eq:DOCoefficients} and \eqref{eq:DOModes} to find the following evolution equation for the mean
\begin{equation}
\frac{\partial \bar{u}}{\partial t} = F_0 + C_{ij} F_{ij} + E[\dot{c}] \frac{\partial \bar{u}}{\partial x} +  E[\dot{c} Y_i] \frac{\partial \hat{u}_i}{\partial x}.
\label{eq:DOMeanKdV}
\end{equation}
We then have the following evolution equation for the stochastic coefficients
\begin{align}
\frac{\mathrm{d}Y_i}{\mathrm{d}t} &= Y_m \, \langle F_m, \hat{u}_i \rangle + (Y_m Y_n - C_{mn}) \langle F_{mn}, \hat{u}_i \rangle \\
&\quad + (\dot{c} - E[\dot{c}]) \langle \frac{\partial \bar{u}}{\partial x}, \hat{u}_i \rangle + (\dot{c} Y_m - E[\dot{c} Y_m]) \langle \frac{\partial \hat{u}_m}{\partial x}, \hat{u}_i \rangle,
\label{eq:DOCoefficientsKdV}
\end{align}
while the modes are governed by the following equation
\begin{equation}
\frac{\partial \hat{u}_i}{\partial t} = \hat{H}_i - \langle \hat{H}_i, \hat{u}_j \rangle \hat{u}_j,
\label{eq:DOModesKdV}
\end{equation}
where the deterministic fields $\hat{H}_i$ are defined as
\begin{align}
\hat{H}_i = F_i + M_{mnk} C_{ik}^{-1} F_{mn} + C_{ik}^{-1} E[\dot{c} Y_k] \frac{\partial \bar{u}}{\partial x} + C_{ik}^{-1} E[\dot{c} Y_k Y_m] \frac{\partial \hat{u}_m}{\partial x},
\label{eq:DOModesHKdV}
\end{align}
with $M_{mnk} = E[Y_m Y_n Y_k]$ the matrix of third-order moments. Finally, the time evolution of the shift amount is given by the stochastic equation \eqref{eq:PhaseParametersDynamicsKdV}, which can be expanded as
\begin{equation}
\dot{c} = \frac{\langle F_0, t' \rangle + Y_i \langle F_i, t' \rangle + Y_i Y_j \langle F_{ij}, t' \rangle}{\langle t(\bar{u}), t' \rangle + Y_i \langle t(\hat{u}_i), t' \rangle}.
\label{eq:PhaseParametersDynamicsKdVSDO}
\end{equation}
At any given time, the full stochastic state can then be reconstructed from the reduced-order quantities and the time-integrated stochastic shift amount through the following group transformation
\begin{equation}
u(x,t;\omega) = g(c(t;\omega)) \hat{u}(x,t;\omega) = \bar{u}(x-c(t;\omega),t) + Y_i(t;\omega) \hat{u}_i(x-c(t;\omega),t).
\label{eq:ReconstructionKdV}
\end{equation}

\subsection{Choice of the template}
\label{sec:ChoiceTemplateKdV}

The SDO equations become singular as the denominator in \eqref{eq:PhaseParametersDynamicsKdVSDO} vanishes. A critical issue is therefore to choose a template which ensures that the quantity $\langle t(\hat{u}), t' \rangle$ always remains nonzero, for all realizations and at all times. For deterministic scalar systems equivariant under continuous translations, this issue was elegantly addressed by \cite{budanur2015} with what they called the \textit{first Fourier mode slice}, a particular choice of template for which the slice condition \eqref{eq:SliceCondition} is equivalent to setting the phase of the first Fourier mode of the symmetry-reduced state $\hat{u}$ to a fixed value. In this way, $\hat{u}$ is pinned at a specific location in physical space and the method is expected to work as long as the amplitude of the first Fourier mode of $u$ does not vanish. 

Consider an arbitrary deterministic state $u$ and its symmetry-reduced counterpart $\hat{u} = g^{-1}(c) u$, where the translation amount $c$ is such that the slice condition $\langle \hat{u}, t(\hat{u}') \rangle = 0$ is satisfied given a fixed template function $\hat{u}'$. Introducing the Fourier series decomposition
\begin{equation}
u(x) = \sum_{k \in \mathbb{Z}} \tilde{u}(k) e^{i2\pi k x/L},
\end{equation}
where $\tilde{u}(k)$, $k \in \mathbb{Z}$ are the Fourier coefficients of $u$, the symmetry-reduced state is expressed as
\begin{equation}
\hat{u}(x) = u(x+c) = \sum_{k \in \mathbb{Z}} |\tilde{u}(k)| e^{i(\arg \tilde{u}(k) + 2\pi k c/L)} e^{i2\pi k x/L},
\end{equation}
where $|\tilde{u}(k)|$ and $\arg \tilde{u}(k)$ are respectively the modulus and phase angle of $\tilde{u}(k)$. The above relation shows that the Fourier coefficients of $\hat{u}$ are those of $u$ rotated by an amount equal to $2\pi k c/L$. In the first Fourier mode slice method, the $c$ is chosen such that the phase angle of the first Fourier mode of $\hat{u}$ is always a fixed value. Here, we choose this value to be $\pi$ so that the symmetry-reduced state appears centered in the domain. This implies $\arg \tilde{u}(1) + 2\pi c/L = \pi$ or equivalently $c = -\arg \tilde{u}(1)L/2\pi + L/2$. As shown in Appendix \ref{app:FirstFourierModeSliceKdVEquation}, this choice of $c$ corresponds to the following template function
\begin{equation}
\hat{u}' = \cos \frac{2 \pi x}{L}.
\label{eq:TemplateKdV}
\end{equation}

The first Fourier mode slice can be readily applied to our stochastic SDO equations by choosing \eqref{eq:TemplateKdV} as our template function. In this way, all symmetry-reduced realizations $\hat{u}(x,t;\omega)$ have their first Fourier mode phase fixed to $\pi$ and are effectively pinned at the center of the physical domain. Furthermore, equation \eqref{eq:PhaseParametersDynamicsKdVSDO} is well posed as long as the amplitude of the first Fourier mode of $u(x,t;\omega)$ remains finite for all realizations and at all times. This is always true unless one of the realizations has spatial periodicity equal to half that of the domain, which is unlikely to happen in practice.

\subsection{Initialization of the SDO quantities}
\label{sec:InitializationSDOKdV}

Given an initial ensemble of realizations $u_0(x;\omega)$, the initial conditions for the quantities involved in the SDO computation are found by a two-step process, where (i) the symmetry-reduced version $\hat{u}_0(x;\omega)$ of each realization is calculated, leading to the initial stochastic translation amount $c_0(\omega)$, and (ii) the initial symmetry-reduced mean $\bar{u}_0(x)$, modes $\hat{u}_{i0}(x)$ and stochastic coefficients $Y_{i0}(\omega)$ are obtained from a truncated Karhunen-Loeve expansion of $\hat{u}_0(x;\omega)$.

\textit{Step 1.} First, the symmetry-reduced counterparts $\hat{u}_0(x;\omega)  = g^{-1}(c_0(\omega)) u_0(x;\omega)$ of the initial realizations $u_0(x;\omega)$ are calculated from
\begin{equation}
\hat{u}_0(x;\omega) = u_0(x+c_0(\omega);\omega) = \sum_{k \in \mathbb{Z}} \tilde{u}_0(k;\omega) e^{i2\pi k c_0(\omega)/L} e^{i2\pi k x/L},
\end{equation}
where $\tilde{u}_0(k;\omega)$ are the Fourier coefficients associated to each realization $u_0(x;\omega)$. To bring the symmetry-reduced state $\hat{u}_0(x;\omega)$ to the first Fourier mode slice, the initial stochastic translation amount $c_0(\omega)$ is chosen such that the phase angle of the first Fourier mode of $\hat{u}_0(x;\omega)$ is equal to $\pi$ for all realizations $\omega$, that is $c_0(\omega) = -\arg \tilde{u}_0(1;\omega)L/2\pi + L/2$.

\textit{Step 2.} The initial symmetry-reduced realizations $\hat{u}_0(x;\omega)$ can then be approximated through a truncated Karhunen-Loeve expansion, giving a set of symmetry-reduced modes and stochastic coefficients from which the SDO modes and coefficients can be initialized. Defining first the initial symmetry-reduced mean $\bar{u}_0(x) = E[\hat{u}_0(x;\omega)]$, the initial symmetry-reduced orthonormal modes $\hat{u}_{i0}(x)$ are then given by the $s$ most energetic eigenfunctions of the following eigenvalue problem
\begin{equation}
\int_0^L R(x,y) \hat{u}_{i0}(x) \, \mathrm{d}x = \lambda_i \hat{u}_{i0}(y), \quad y \in [0,L], \quad i = 1,...,s,
\label{eq:TruncatedKLInitialConditionKdV}
\end{equation}
where the correlation operator $R(x,y) = E[(\hat{u}_0(x;\omega)-\bar{u}_0(x))(\hat{u}_0(y;\omega)-\bar{u}_0(y))]$. Finally, the associated initial stochastic coefficients $Y_{i0}(\omega)$ are obtained by projection of the initial symmetry-reduced realizations to the symmetry-reduced modes
\begin{equation}
Y_{i0}(\omega) = \langle \hat{u}_0(x;\omega)-\bar{u}_0(x), \hat{u}_{i0}(x) \rangle, \quad i = 1,...,s.
\end{equation}

Note that $s$ not only represents the number of modes and stochastic coefficients of the SDO scheme, but also sets the tolerance on the truncated Karhunen-Loeve decomposition \eqref{eq:TruncatedKLInitialConditionKdV} of the initial condition. Therefore, its value should be chosen so that both (i) the stochastic symmetry-reduced initial condition $\hat{u}_0(x;\omega)$ and (ii) the subsequent time-evolving stochastic solution $\hat{u}(x,t;\omega)$ are sufficiently well approximated. The accuracy of the initial condition can be monitored by the decay of the eigenvalues $\lambda_i$ of the truncated Karhunen-Loeve decomposition \eqref{eq:TruncatedKLInitialConditionKdV}, which indicate the amount of variance in $\hat{u}_0(x;\omega)$ contained along each of the corresponding eigendirections $\hat{u}_{i0}(x)$. Similarly, one can get a rough idea on the accuracy of the stochastic solution by looking at the evolution of the variance in the stochastic coefficients associated with the last few modes. A rigorous approach, however, requires knowledge of the exact solution (either from Monte-Carlo simulations or analytical arguments).

In section \ref{sec:ApplicationStochasticSoliton}, we will compare numerical results from the SDO framework with those from a DO computation. In the case of DO, the various quantities are initialized following Step 2 directly, i.e.~the initial mean $\bar{u}_0(x)$, modes $u_{i0}(x)$ and stochastic coefficients $Y_{i0}(\omega)$ are defined from a truncated Karhunen-Loeve expansion of the initial full state-space realizations $u_0(x;\omega)$.

\subsection{Numerical scheme}

The SDO equations for the mean and the modes are implemented using a standard pseudo-spectral method in space with 3/2 antialiasing and a semi-implicit Euler scheme in time, where the third-order derivatives are treated implicitly and the nonlinear terms are treated explicitly. The stochastic coefficients and translation amount are integrated in time with respectively a 4th-order Runge-Kutta and a 2-step Adams-Bashforth scheme, both using a particle method. Note that setting the stochastic translation amount to zero in the SDO equations readily yields the standard DO scheme. For both SDO and DO, we use $L = 2\pi$, $\mu = 5 \cdot 10^{-4}$, 512 Fourier modes, $\Delta t = 10^{-4}$ and 1000 Monte-Carlo particles.

\subsection{Example with a stochastic soliton}
\label{sec:ApplicationStochasticSoliton}

The KdV equation admits a class of solitary wave solutions with shape-dependent propagation speed. In this section, we use the SDO equations to compute the evolution of a stochastic initial condition consisting of such solitons
\begin{equation}
u_0(x;\omega) = 3a(\omega) \, \mathrm{sech}^2 \left[ \sqrt{\frac{a(\omega)}{\mu}} \frac{x-L/2}{2} \right],
\label{eq:SolitonKdV}
\end{equation}
where $a(\omega) \sim \mathcal{U}(0.1,0.5)$ is a random variable that affects both the amplitude and the width of the hyperbolic secant profile. Each realization $\omega$ defined by the above initial condition is an exact soliton solution of the KdV equation with propagation velocity $a(\omega)$, which makes this problem challenging for classical order reduction methods as a large number of modes eventually becomes necessary to reproduce faithfully the spatial dispersion of all realisations. The SDO scheme, on the other hand, should not suffer from such issues since the shift amount $c(t;\omega)$ takes care of the spatial translation of the realizations, leaving the symmetry-reduced mean and modes to account for a change in shape that is here nonexistent since the solution consists of solitons. Hence, we use just one mode for the SDO computation, which we initialize according to the procedure described in Section \ref{sec:InitializationSDOKdV}. For comparison purposes, we also perform a regular DO simulation of the same problem, but using 10 modes instead. 

\begin{figure}
\centering
\includegraphics[width=0.49\textwidth]{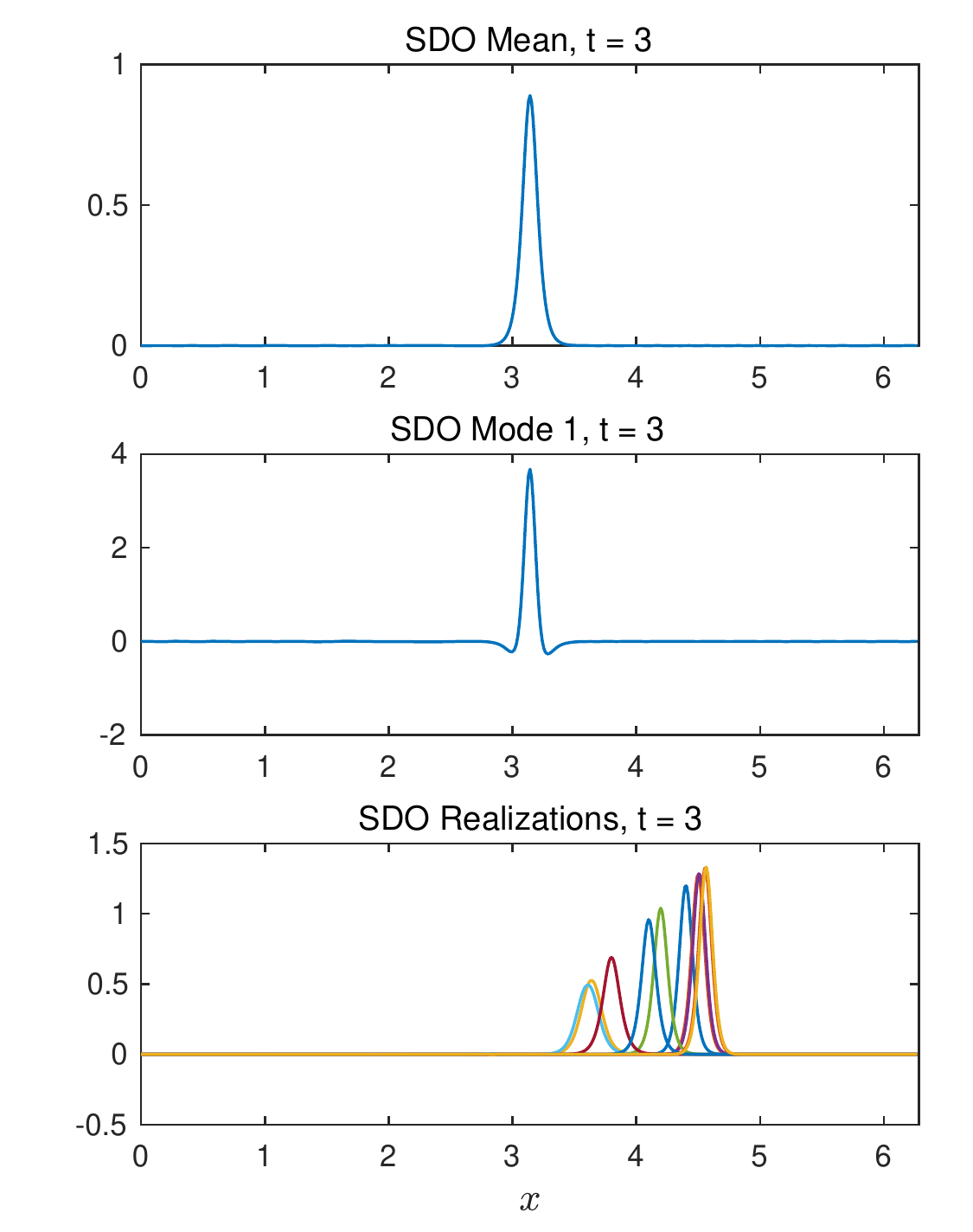}
\includegraphics[width=0.49\textwidth]{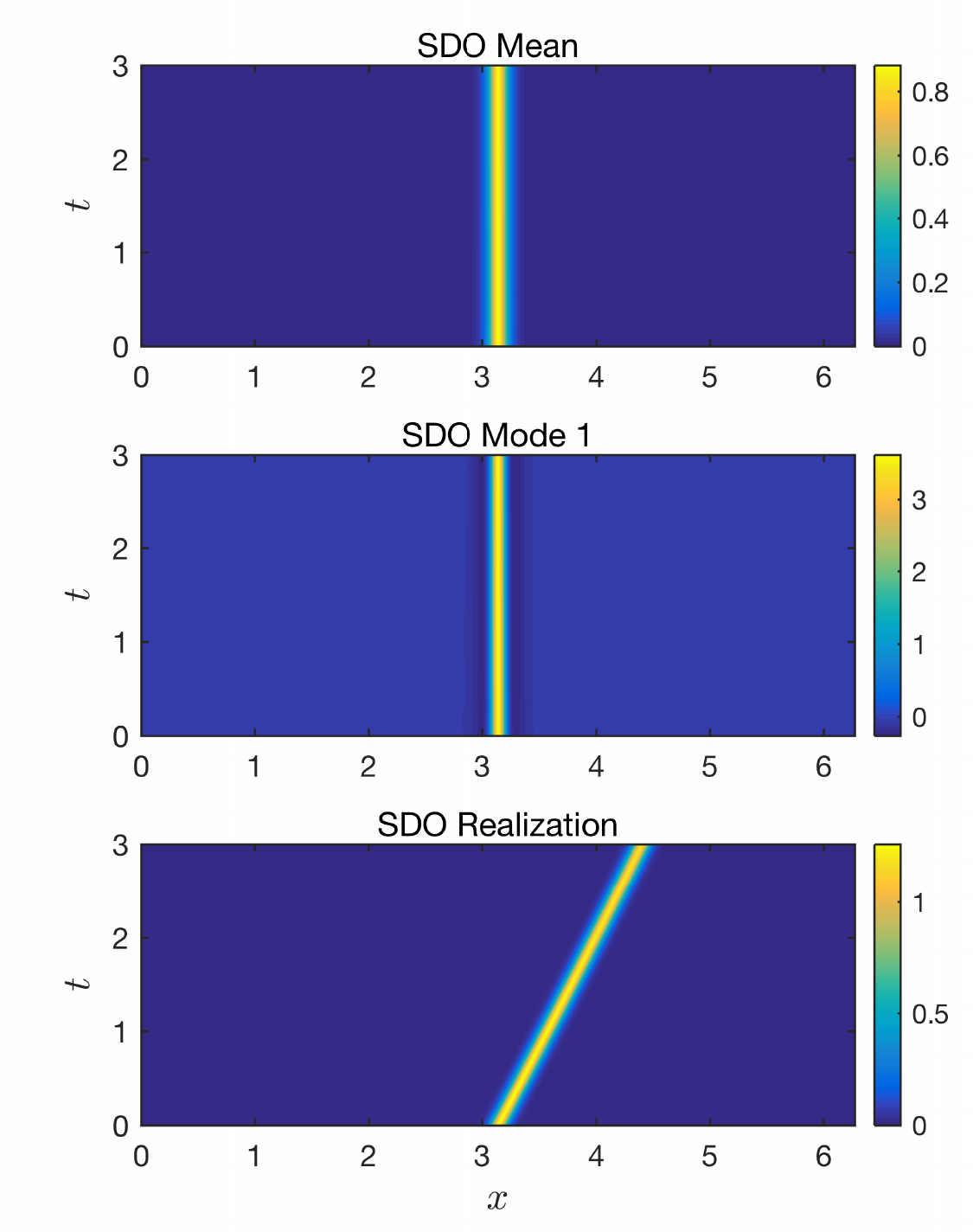}
\caption{Symmetry-reduced mean $\bar{u}$ and mode $\hat{u}_1$ of the SDO solution with 1 mode at final time $t = 3$ (left) and spatio-temporal evolution (right). We also show 10 full state-space realizations at final time (left) and the spatio-temporal evolution of one of those realizations (right). The full state-space realizations are recovered from their symmetry-reduced state as $u = g(c) \hat{u}$.}
\label{fig:SDOKdV}
\end{figure}

The first two rows in Figure \ref{fig:SDOKdV} show the symmetry-reduced mean $\bar{u}(x,t)$ and the single symmetry-reduced mode $\hat{u}_1(x,t)$ of the SDO solution at final time $t = 3$ (left), and their spatio-temporal evolution (right). 
As expected, both the symmetry-reduced mean and mode are steady, which reflects the fact that the dynamics of the individual solitons purely consist of translation, which is entirely absorbed in the stochastic shift amount $c(t;\omega)$. The original stochastic state $u(x,t;\omega)$ can be reconstructed from the SDO quantities through the group transformation \eqref{eq:ReconstructionKdV}, and we display in the last row of Figure \ref{fig:SDOKdV} ten full state-space realizations at final time $t = 3$ (left) and the spatio-temporal evolution of one of them (right). Each individual soliton realization maintains its initial shape while propagating at a constant amplitude-dependent velocity, which is expected but nonetheless remarkable considering that the SDO computation uses just one single mode. 

\begin{figure}
\centering
\includegraphics[width=0.49\textwidth]{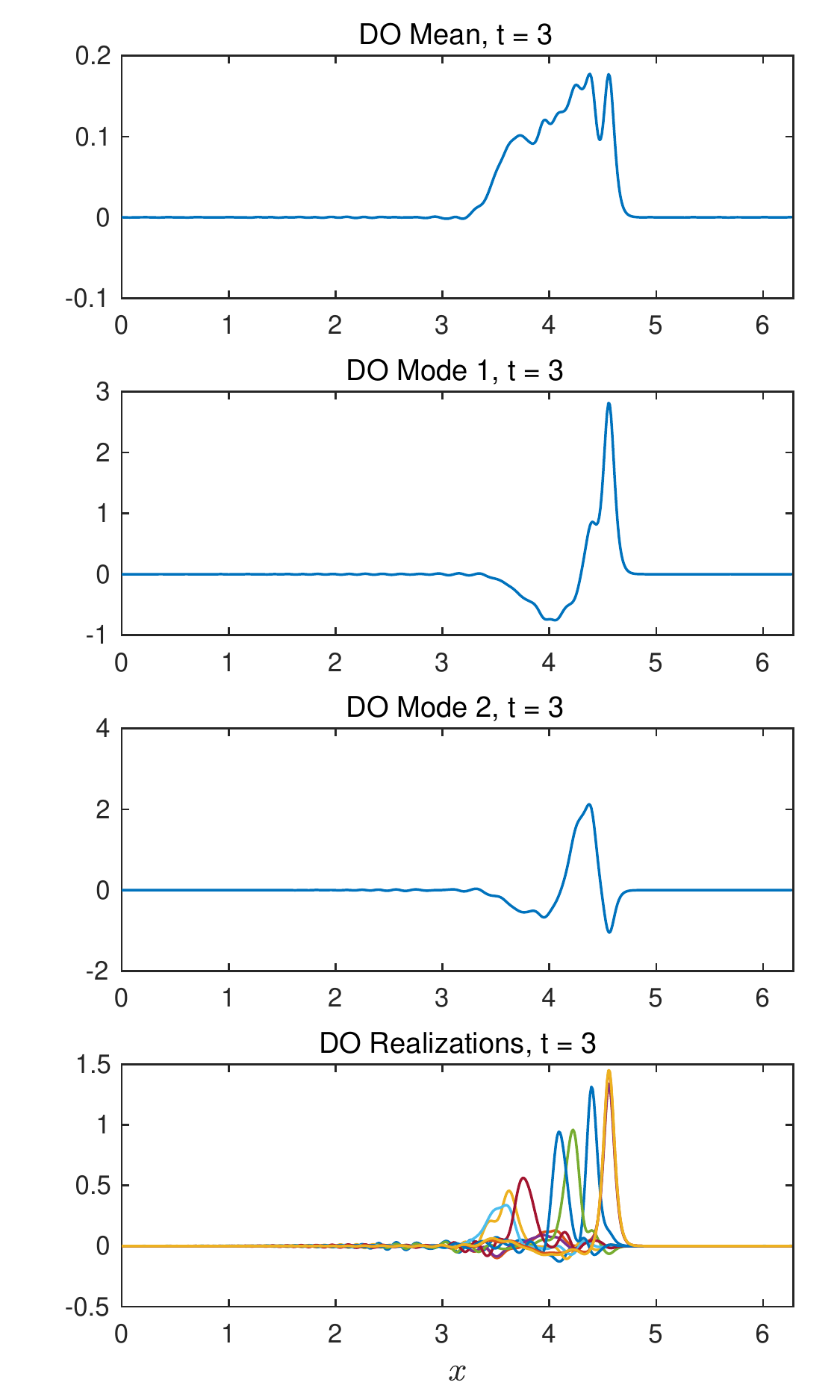}
\includegraphics[width=0.49\textwidth]{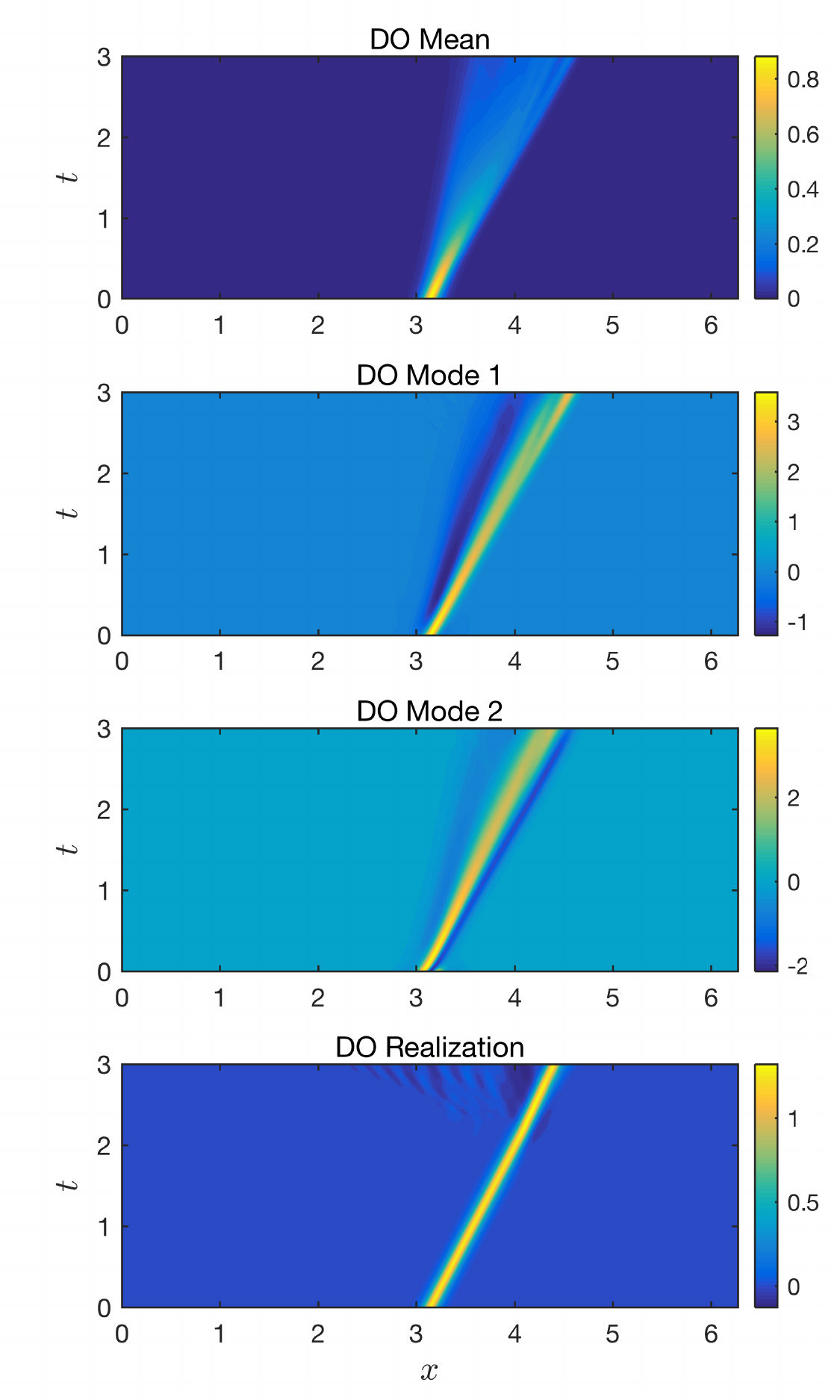}
\caption{Mean $\bar{u}$ and modes $u_i$ of the DO solution with 10 modes at final time $t = 3$ (left) and spatio-temporal evolution (right). We also show 10 realizations at final time (left) and the spatio-temporal evolution of one of those realizations (right). In the case of DO, the mean, modes and realization already correspond to the full state-space solution.}
\label{fig:DOKdV}
\end{figure}

To put these results into context, Figure \ref{fig:DOKdV} shows the mean $\bar{u}(x,t)$, the first two modes $u_i(x,t)$ and ten full state-space realizations $u(x,t;\omega)$ of the corresponding DO simulation.
Contrary to the SDO simulation, the DO realizations at final time have not preserved their initial shape although the DO simulation is using ten times as many modes. This is because the DO mean and modes need to account for the translation at random speed of the stochastic soliton, hence they become dispersed over space as time progresses. 

\begin{figure}
\centering
\includegraphics[width=0.49\textwidth]{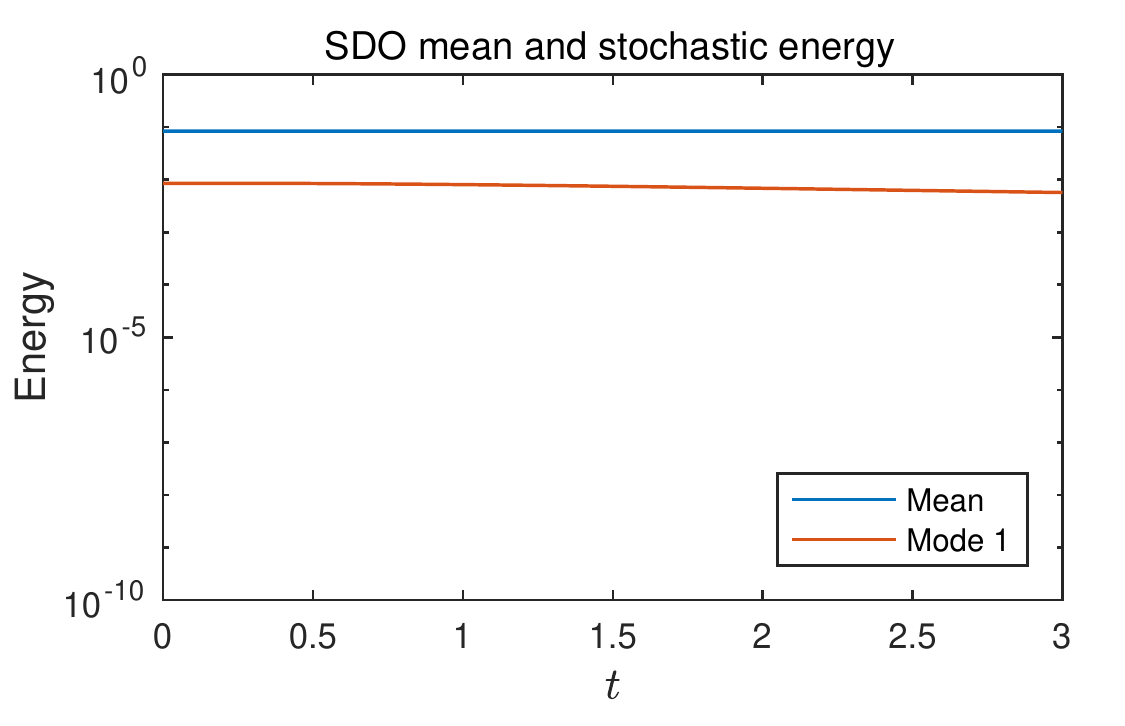}
\includegraphics[width=0.49\textwidth]{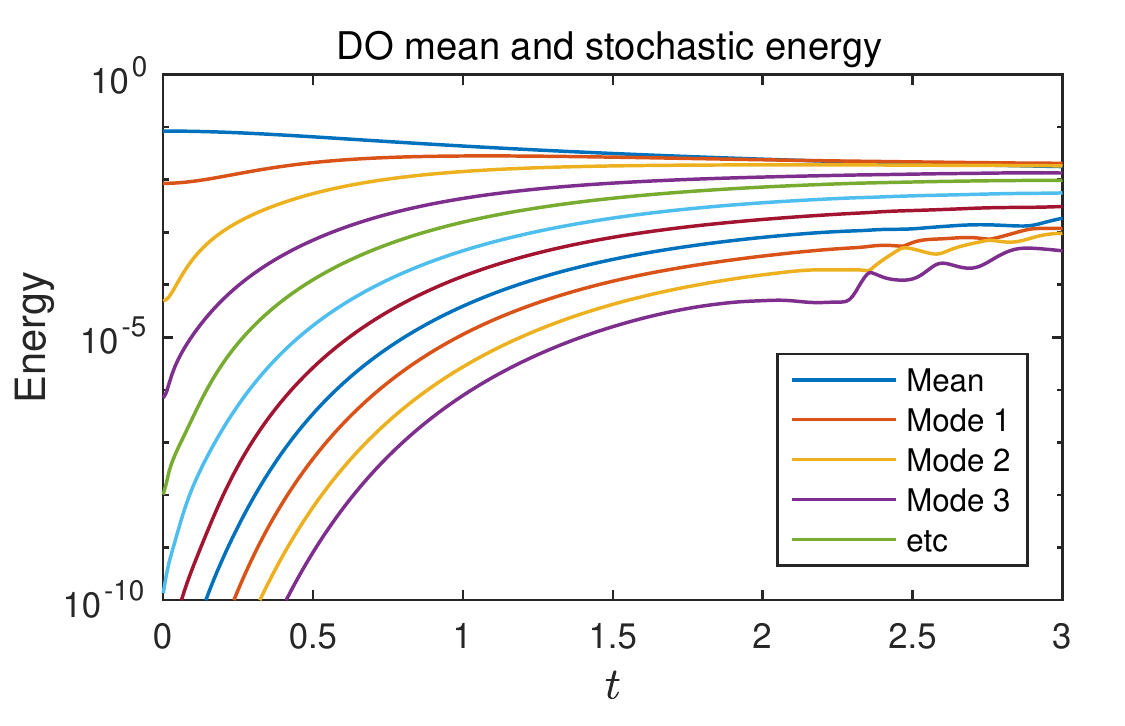}
\caption{Energy in the mean $\langle \bar{u},\bar{u} \rangle$ and variance of the stochastic coefficients $E[Y_i^2]$ of the SDO solution with 1 mode (left) and DO solution with 10 modes (right).}
\label{fig:VarianceKdV}
\end{figure}

This difference in the amount of stochasticity accounted for by the modes is clearly observed in Figure \ref{fig:VarianceKdV}, where we plot the time evolution of the energy in the mean $\langle \bar{u},\bar{u} \rangle$ and the variance of the stochastic coefficients $E[Y_i^2]$ for the SDO (left) and DO (right) solutions.
The stochastic energy of the SDO mode, represented by the variance of the corresponding stochastic coefficient, remains equal to its initial value while that of the DO modes grows drastically over time. Thus, the DO scheme would need a much larger number of modes to achieve a comparable level of accuracy to the SDO simulation, which proves the better efficiency of SDO.

\section{Application to the Navier-Stokes equations}
\label{sec:ApplicationNavierStokesEquations}

In this section, we illustrate our approach with the two-dimensional Navier-Stokes equations
\begin{subequations}
\begin{gather}
\frac{\partial \mathbf{u}}{\partial t} + \mathbf{u} \cdot \nabla \mathbf{u} = - \nabla p + \frac{1}{Re} \Delta \mathbf{u}, \\
\nabla \cdot \mathbf{u} = 0,
\end{gather}
\end{subequations}
written in nondimensional variables, with $Re$ the Reynolds number, and defined on the spatial domain $\mathbf{x} \in D = [0,L_1] \times [0,L_2]$ with periodic boundary conditions. The Navier-Stokes equations are equivariant under translations, rotations and inversion about the origin. Focusing on the group of continuous translations along the two spatial directions $G = \mathrm{SO}(2)_{x_1} \times \mathrm{SO}(2)_{x_2}$, we define the shift operator $g \in G$ as
\begin{equation}
g(\mathbf{c})\mathbf{u}(\mathbf{x}) = \mathbf{u}(\mathbf{x}-\mathbf{c}),
\end{equation}
where the continuous phase parameter $\mathbf{c} = (c_1,c_2)^\mathsf{T}$ represents the shift amounts in the $x_1$ and $x_2$ directions. The group orbit $g\mathbf{u}$ at an arbitrary state $\mathbf{u}$ thus possesses two tangents, one for each direction $x_1$ and $x_2$, given by
\begin{subequations}
\begin{gather}
\mathbf{t}_1(\mathbf{u}) = \lim_{\delta c_1 \rightarrow 0} \frac{g(\delta c_1,0)\mathbf{u} - \mathbf{u}}{\delta c_1} = - \frac{\partial \mathbf{u}}{\partial x_1}, \\
\mathbf{t}_2(\mathbf{u}) = \lim_{\delta c_2 \rightarrow 0} \frac{g(0,\delta c_2)\mathbf{u} - \mathbf{u}}{\delta c_2} = - \frac{\partial \mathbf{u}}{\partial x_2}.
\end{gather}
\end{subequations}

\subsection{Dynamics within the symmetry-reduced state space} 

As before, we now pick a fixed template $\hat{\mathbf{u}}'(\mathbf{x})$ and we consider the evolution of the symmetry-reduced state $\hat{\mathbf{u}}(\mathbf{x},t;\omega) = g^{-1}(\mathbf{c}(t;\omega)) \mathbf{u}(\mathbf{x},t;\omega)$ such that the distance $||\hat{\mathbf{u}}(\mathbf{x},t;\omega) - \hat{\mathbf{u}}'(\mathbf{x})||$ between the reduced state and the template is minimized at all times $t$ and for all realizations $\omega$. In the case of the Navier-Stokes equations subject to the symmetry group $G$, the dynamical equations \eqref{eq:ReducedStateDynamics} for $\hat{\mathbf{u}}$ in the symmetry-reduced state space become
\begin{equation}
\frac{\partial \hat{\mathbf{u}}}{\partial t} = \mathbf{F}(\hat{\mathbf{u}}) + \dot{\mathbf{c}} \cdot \nabla \hat{\mathbf{u}},
\label{eq:ReducedStateDynamicsNS}
\end{equation}
where the full state space nonlinear differential operator $\mathbf{F}$ has the form
\begin{equation}
\mathbf{F}(\hat{\mathbf{u}}) = - \nabla p + \frac{1}{Re} \Delta \hat{\mathbf{u}} - \hat{\mathbf{u}} \cdot \nabla \hat{\mathbf{u}},
\end{equation}
with $p$ acting as a Lagrange multiplier to enforce $\nabla \cdot \hat{\mathbf{u}} = 0$ since the divergence-free condition on $\mathbf{u}$ implies that $\hat{\mathbf{u}}$ is also divergence-free. Finally, the stochastic differential equation \eqref{eq:PhaseParametersDynamics} for the shift amount $\mathbf{c}(t;\omega)$ writes 
\begin{equation}
\dot{\mathbf{c}} = \mathbf{T}^{-1} \mathbf{f} =
\left[
\begin{array}{cc}
\langle \mathbf{t}_1(\hat{\mathbf{u}}), \mathbf{t}_1' \rangle & \langle \mathbf{t}_2(\hat{\mathbf{u}}), \mathbf{t}_1' \rangle \\
\langle \mathbf{t}_1(\hat{\mathbf{u}}), \mathbf{t}_2' \rangle & \langle \mathbf{t}_2(\hat{\mathbf{u}}), \mathbf{t}_2' \rangle
\end{array}
\right]^{-1}
\left[
\begin{array}{c}
\langle \mathbf{F}(\hat{\mathbf{u}}), \mathbf{t}_1' \rangle \\
\langle \mathbf{F}(\hat{\mathbf{u}}), \mathbf{t}_2' \rangle
\end{array}
\right].
\label{eq:PhaseParametersDynamicsNS}
\end{equation}

\subsection{Order reduction in the symmetry-reduced state space}

In order to formulate the reduced-order SDO equations, let us first approximate the symmetry-reduced stochastic state as a truncated Karhunen-Loeve expansion
\begin{equation}
\hat{\mathbf{u}}(\mathbf{x},t;\omega) = \bar{\mathbf{u}}(\mathbf{x},t) + Y_i(t;\omega) \hat{\mathbf{u}}_i(\mathbf{x},t), \quad i = 1,...,s,
\end{equation}
where $\bar{\mathbf{u}}$ and $\hat{\mathbf{u}}_i$ are respectively the symmetry-reduced mean and modes and $Y_i$ are the stochastic coefficients of the finite-dimensional expansion. Next, we insert the above representation into the RHS of the evolution equation \eqref{eq:ReducedStateDynamicsNS} for the symmetry-reduced state to obtain
\begin{equation}
\mathbf{F}(\hat{\mathbf{u}}) + \dot{\mathbf{c}} \cdot \nabla \hat{\mathbf{u}} = - \nabla p + \mathbf{F}_0 + Y_i \, \mathbf{F}_i + Y_i Y_j \, \mathbf{F}_{ij} + \dot{\mathbf{c}} \cdot \nabla \bar{\mathbf{u}} + \dot{\mathbf{c}} Y_i \cdot \nabla \hat{\mathbf{u}}_i,
\label{eq:ExpandedOperatorNS}
\end{equation}
where $\mathbf{F}_0$, $\mathbf{F}_i$ and $\mathbf{F}_{ij}$ are deterministic fields given by
\begin{subequations}
\begin{gather}
\mathbf{F}_0 = \frac{1}{Re} \Delta \bar{\mathbf{u}} - \bar{\mathbf{u}} \cdot \nabla \bar{\mathbf{u}}, \\
\mathbf{F}_i = \frac{1}{Re} \Delta \hat{\mathbf{u}}_i - \hat{\mathbf{u}}_i \cdot \nabla \bar{\mathbf{u}} - \bar{\mathbf{u}} \cdot \nabla \hat{\mathbf{u}}_i, \\
\mathbf{F}_{ij} = - \hat{\mathbf{u}}_i \cdot \nabla \hat{\mathbf{u}}_j.
\end{gather}
\end{subequations}
The stochastic pressure $p$ ensures that $\hat{\mathbf{u}}$ is divergence-free, hence by setting the divergence of the expanded RHS operator \eqref{eq:ExpandedOperatorNS} to zero (following \cite{sapsis2013}), one finds that
\begin{equation}
p = p_0 + Y_i \, p_i + Y_i Y_j \, p_{ij},
\end{equation}
where $\Delta p_0 = \nabla \cdot \mathbf{F}_0$, $\Delta p_i = \nabla \cdot \mathbf{F}_i$ and $\Delta p_{ij} = \nabla \cdot \mathbf{F}_{ij}$ are deterministic pressure fields. Now, we insert the expanded RHS operator \eqref{eq:ExpandedOperatorNS} governing the evolution of the symmetry-reduced state $\hat{\mathbf{u}}$ into the SDO equations \eqref{eq:DOMean}, \eqref{eq:DOCoefficients} and \eqref{eq:DOModes} to find the following evolution equation for the mean
\begin{equation}
\frac{\partial \bar{\mathbf{u}}}{\partial t} = -\nabla p_0 + \mathbf{F}_0 + C_{ij} (-\nabla p_{ij} + \mathbf{F}_{ij}) + E[\dot{\mathbf{c}}] \cdot \nabla \bar{\mathbf{u}} +  E[\dot{\mathbf{c}} Y_i] \cdot \nabla \hat{\mathbf{u}}_i.
\label{eq:DOMeanNS}
\end{equation}
We then have the following evolution equation for the stochastic coefficients
\begin{align}
\frac{\mathrm{d} Y_i}{\mathrm{d} t} &= Y_m \langle - \nabla p_m + \mathbf{F}_m, \hat{\mathbf{u}}_i \rangle + (Y_m Y_n - C_{mn}) \langle - \nabla p_{mn} + \mathbf{F}_{mn}, \hat{\mathbf{u}}_i \rangle \nonumber \\
&\quad + (\dot{\mathbf{c}} - E[\dot{\mathbf{c}}]) \cdot \langle \nabla \bar{\mathbf{u}}, \hat{\mathbf{u}}_i \rangle + (\dot{\mathbf{c}} Y_m - E[\dot{\mathbf{c}} Y_m]) \cdot \langle \nabla \hat{\mathbf{u}}_m, \hat{\mathbf{u}}_i \rangle,
\label{eq:DOCoefficientsNS}
\end{align}
where we have defined $\langle \nabla \bar{\mathbf{u}}, \hat{\mathbf{u}}_i \rangle = (\langle \partial_{x_1} \bar{\mathbf{u}}, \hat{\mathbf{u}}_i \rangle, \langle \partial_{x_2} \bar{\mathbf{u}}, \hat{\mathbf{u}}_i \rangle)^\mathsf{T}$, and similarly $\langle \nabla \hat{\mathbf{u}}_m, \hat{\mathbf{u}}_i \rangle = (\langle \partial_{x_1} \hat{\mathbf{u}}_m, \hat{\mathbf{u}}_i \rangle, \langle \partial_{x_2} \hat{\mathbf{u}}_m, \hat{\mathbf{u}}_i \rangle)^\mathsf{T}$. The modes are governed by the following equation
\begin{equation}
\frac{\partial \hat{\mathbf{u}}_i}{\partial t} = \hat{\mathbf{H}}_i - \langle \hat{\mathbf{H}}_i, \hat{\mathbf{u}}_j \rangle \hat{\mathbf{u}}_j,
\label{eq:DOModesNS}
\end{equation}
where the deterministic fields $\hat{\mathbf{H}}_i$ are defined as
\begin{align}
\hat{\mathbf{H}}_i &= -\nabla p_i + \mathbf{F}_i + M_{kmn} C_{ik}^{-1} (-\nabla p_{mn} + \mathbf{F}_{mn}) \nonumber \\
&\quad + C_{ik}^{-1} E[\dot{\mathbf{c}} Y_k] \cdot \nabla \bar{\mathbf{u}} + C_{ik}^{-1} E[\dot{\mathbf{c}} Y_k Y_m] \cdot \nabla \hat{\mathbf{u}}_m.
\label{eq:DOModesHNS}
\end{align}
with $M_{kmn} = E[Y_k Y_m Y_n]$ the matrix of third-order moments. Finally, the time evolution of the shift amount $\dot{\mathbf{c}}$ is given by the stochastic linear system \eqref{eq:PhaseParametersDynamicsNS} where the stochastic matrix $\mathbf{T}$ and vector $\mathbf{f}$ can be expanded in functions of $Y_i$ and $Y_i Y_j$ with $i,j = 1,...,s$ using that $\mathbf{t}_a(\hat{\mathbf{u}}) = \mathbf{t}_a(\bar{\mathbf{u}}) + Y_i \mathbf{t}_a(\hat{\mathbf{u}}_i)$, $a = 1,2$ and $\mathbf{F}(\hat{\mathbf{u}}) = - \nabla p_0 + \mathbf{F}_0 + Y_i \, (- \nabla p_i + \mathbf{F}_i) + Y_i Y_j \, (- \nabla p_{ij} + \mathbf{F}_{ij})$.

At any given time, the full stochastic state can be reconstructed from the reduced-order quantities and the time-integrated stochastic shift amount through the group transformation
\begin{equation}
\mathbf{u}(\mathbf{x},t;\omega) = g(\mathbf{c}(t;\omega)) \hat{\mathbf{u}}(\mathbf{x},t;\omega) = \bar{\mathbf{u}}(\mathbf{x}-\mathbf{c}(t;\omega),t) + Y_i(t;\omega) \hat{\mathbf{u}}_i(\mathbf{x}-\mathbf{c}(t;\omega),t;\omega).
\label{eq:ReconstructionNS}
\end{equation}

\subsection{Choice of the template}
\label{sec:ChoiceTemplateNS}

The first Fourier mode slice is unambiguously defined for scalar fields since there is a unique Fourier coefficient with wavenumber one which defines the symmetry-reduced state. However, ambiguity arises when one considers vector fields with multiples components and therefore multiple Fourier coefficients of wavenumber one. Nevertheless, we show here that for the particular case of fluid flow in two dimensions, one can overcome this issue by defining the first Fourier mode slice based on the scalar vorticity field instead.

Let us introduce an arbitrary deterministic two-dimensional velocity field $\mathbf{u}(\mathbf{x}) = (u_1(x_1,x_2),u_2(x_2,x_2))^\mathsf{T}$ and its symmetry-reduced counterpart $\hat{\mathbf{u}} = g^{-1}(\mathbf{c}) \mathbf{u}$, where $\mathbf{c} = (c_1,c_2)^\mathsf{T}$ is such that both slice conditions $\langle \hat{\mathbf{u}},\mathbf{t}_1(\hat{\mathbf{u}}') \rangle = 0$ and $\langle \hat{\mathbf{u}},\mathbf{t}_2(\hat{\mathbf{u}}') \rangle = 0$ are satisfied given a fixed template function $\hat{\mathbf{u}}'$. Consider now the scalar vorticity field defined by $\omega = \partial_{x_1}u_2 - \partial_{x_2}u_1$. Its Fourier series is
\begin{equation}
\omega(\mathbf{x}) = \sum_{k_1, k_2 \in \mathbb{Z}} \tilde{\omega}(k_1,k_2) e^{i 2\pi (k_1 x_1/L_1+k_2 x_2/L_2)},
\end{equation}
where $\tilde{\omega} = i k_1 \tilde{u}_2 - i k_2 \tilde{u}_1$, with $\tilde{u}_1$ and $\tilde{u}_2$ the Fourier coefficients of $u_1$ and $u_2$ respectively. The corresponding symmetry-reduced vorticity field is expressed as
\begin{align}
\hat{\omega}(\mathbf{x}) &= \omega(\mathbf{x}+\mathbf{c}) \nonumber \\
&= \sum_{k_1, k_2 \in \mathbb{Z}} |\tilde{\omega}(k_1,k_2)| e^{i(\arg \tilde{\omega}(k_1,k_2) + 2\pi (k_1 c_1/L_1 + k_2 c_2/L_2))} e^{i 2\pi (k_1 x_1/L_1+k_2 x_2/L_2)},
\label{eq:SymmetryReducedStateNS}
\end{align}
where $|\tilde{\omega}(k_1,k_2)|$ and $\arg \tilde{\omega}(k_1,k_2)$ are respectively the modulus and phase angle of $\tilde{\omega}$. Similar to the one-dimensional case, the above relation shows that the Fourier coefficients of $\hat{\omega}$ are those of $\omega$ rotated by an amount equal to $2\pi (k_1 c_1/L_1 + k_2 c_2/L_2)$. Therefore we can define the first Fourier mode slice by choosing $c_1$ and $c_2$ such that the phase angles of the first Fourier modes of $\hat{\omega}$ in the $x_1$ and $x_2$ directions are equal to $\pi$, i.e.~$\arg \tilde{\omega}(1,0) + 2\pi c_1/L_1 = \pi$ and $\arg \tilde{\omega}(0,1) + 2\pi c_2/L_2 = \pi$. In this way, the symmetry-reduced vorticity field will appear centered in the domain. We can now use $\arg \tilde{\omega}(1,0) = \arg \tilde{u}_2(1,0) + \pi/2$ and $\arg \tilde{\omega}(0,1) = \arg \tilde{u}_1(0,1) - \pi/2$ to express this choice of $c_1$ and $c_2$ in terms of the Fourier coefficients of the velocity field, resulting in $c_1 = -\arg \tilde{u}_2(1,0) L_1/2\pi + L_1/4$ and $c_2 = -\arg \tilde{u}_1(0,1) L_2/2\pi - L_2/4$. Finally, we show in Appendix \ref{app:FirstFourierModeSliceNSEquation} that this choice of $\mathbf{c}$ corresponds to the template function
\begin{equation}
\hat{\mathbf{u}}' = \left( \sin \frac{2 \pi x_2}{L_2}, \sin \frac{2 \pi x_1}{L_1} \right)^\mathsf{T}.
\label{eq:TemplateNS}
\end{equation}

The first Fourier mode slice defined above can be readily applied to our stochastic SDO equations by choosing \eqref{eq:TemplateNS} as our template function. In this way, all symmetry-reduced realizations $\hat{\mathbf{u}}(\mathbf{x},t;\omega)$ have the phase of their first Fourier modes in the $x$ and $y$ directions fixed to $\pi$ and therefore remain centered in the two-dimensional physical space. Furthermore, it is possible to show that using template \eqref{eq:TemplateNS}, the matrix $\mathbf{T}$ appearing in equation \eqref{eq:PhaseParametersDynamicsNS} is diagonal, with the two diagonal elements given respectively by the amplitudes of the first Fourier modes of $u_2$ in the $x$ direction and $u_1$ in the $y$ direction. As such, $\mathbf{T}$ is well conditioned so long as the ratio of these two quantities does not get large, and equation \eqref{eq:PhaseParametersDynamicsNS} is well posed as long as these quantities remain nonzero for all realizations and at all times. This condition is always true unless one of the realizations has spatial periodicity equal to half that of the domain, which is unlikely to happen in practice.

\subsection{Initialization of the SDO quantities}
\label{sec:InitializationSDONS}

Given an initial ensemble of realizations $\mathbf{u}_0(\mathbf{x};\omega)$, the initial conditions for the quantities involved in the SDO computation are found by a two-step process similar to Section \ref{sec:InitializationSDOKdV}, where (i) the symmetry-reduced version $\hat{\mathbf{u}}_0(\mathbf{x};\omega)$ of each realization is calculated, leading to the initial stochastic translation amount $\mathbf{c}_0(\omega)$, and (ii) the initial symmetry-reduced mean $\bar{\mathbf{u}}_0(\mathbf{x})$, modes $\hat{\mathbf{u}}_{i0}(\mathbf{x})$ and stochastic coefficients $Y_{i0}(\omega)$ are obtained from a truncated Karhunen-Loeve expansion of $\hat{\mathbf{u}}_0(\mathbf{x};\omega)$.

\textit{Step 1.} First, the symmetry-reduced counterparts $\hat{\mathbf{u}}_0(\mathbf{x};\omega)  = g^{-1}(\mathbf{c}_0(\omega)) u_0(\mathbf{x};\omega)$ of the initial realizations $\mathbf{u}_0(\mathbf{x};\omega)$ are given by
\begin{align}
\hat{\mathbf{u}}_0(\mathbf{x};\omega) &= \mathbf{u}_0(\mathbf{x}+\mathbf{c}_0(\omega);\omega) \nonumber \\
&= \sum_{k_1,k_2 \in \mathbb{Z}} \tilde{\mathbf{u}}_0(k_1,k_2;\omega) e^{i2\pi (k_1 c_{10}(\omega)/L_1 + k_2 c_{20}(\omega)/L_2)} e^{i2\pi (k_1 x_1/L_1 + k_2 x_2/L_2)},
\end{align}
where $\tilde{\mathbf{u}}_0(k_1,k_2;\omega) = (\tilde{u}_{10}(k_1,k_2;\omega),\tilde{u}_{20}(k_1,k_2;\omega))^\mathsf{T}$ are the Fourier coefficients associated to each realization $\mathbf{u}_0(\mathbf{x};\omega) = (u_{10}(\mathbf{x};\omega),u_{20}(\mathbf{x};\omega))^\mathsf{T}$. To bring the symmetry-reduced state $\hat{\mathbf{u}}_0(\mathbf{x};\omega)$ to the first Fourier mode slice, we proceed according to Section \ref{sec:ChoiceTemplateNS} and choose the initial stochastic translation amount $\mathbf{c}_0(\omega) = (c_{10}(\omega),c_{20}(\omega))^\mathsf{T}$ as $c_{10}(\omega) = -\arg \tilde{u}_{20}(1,0;\omega) L_1/2\pi + L_1/4$ and $c_{20}(\omega) = -\arg \tilde{u}_{10}(0,1;\omega) L_2/2\pi - L_2/4$.

\textit{Step 2.} The initial symmetry-reduced realizations $\hat{\mathbf{u}}_0(\mathbf{x};\omega)$ can then be approximated through a truncated Karhunen-Loeve expansion, giving a set of symmetry-reduced modes and stochastic coefficients from which the SDO modes and coefficients can be initialized. Defining first the initial symmetry-reduced mean $\bar{\mathbf{u}}_0(\mathbf{x}) = E[\hat{\mathbf{u}}_0(\mathbf{x};\omega)]$, the initial symmetry-reduced orthonormal modes $\hat{\mathbf{u}}_{i0}(\mathbf{x})$ are then given by the $s$ most energetic eigenfunctions of the following eigenvalue problem
\begin{equation}
\iint_D \mathbf{R}(\mathbf{x},\mathbf{y}) \hat{\mathbf{u}}_{i0}(\mathbf{x}) \, \mathrm{d}\mathbf{x} \mathrm{d}\mathbf{y} = \lambda_i \hat{\mathbf{u}}_{i0}(\mathbf{y}), \quad \mathbf{y} \in D, \quad i = 1,...,s,
\end{equation}
where the correlation matrix is $\mathbf{R}(\mathbf{x},\mathbf{y}) = E[(\hat{\mathbf{u}}_0(\mathbf{x};\omega)-\bar{\mathbf{u}}_0(\mathbf{x}))(\hat{\mathbf{u}}_0(\mathbf{y};\omega)-\bar{\mathbf{u}}_0(\mathbf{y}))^\mathsf{T}]$. Finally, the associated initial stochastic coefficients $Y_{i0}(\omega)$ are obtained by projection of the initial symmetry-reduced realizations to the symmetry-reduced modes
\begin{equation}
Y_{i0}(\omega) = \langle \hat{\mathbf{u}}_0(\mathbf{x};\omega)-\bar{\mathbf{u}}_0(\mathbf{x}), \hat{\mathbf{u}}_{i0}(\mathbf{x}) \rangle, \quad i = 1,...,s.
\end{equation}

In section \ref{sec:ExampleVortex}, we will compare numerical results from the SDO framework with results from the regular DO equations. In the case of DO, the various quantities are initialized following Step 2 directly, i.e.~the initial mean $\bar{\mathbf{u}}_0(\mathbf{x})$, modes $\mathbf{u}_{i0}(\mathbf{x})$ and stochastic coefficients $Y_{i0}(\omega)$ are defined from a truncated Karhunen-Loeve expansion of the initial full state space realizations $\mathbf{u}_0(\mathbf{x};\omega)$.

\subsection{Numerical scheme}

The SDO equations for the mean and the modes are implemented using a standard pseudo-spectral method in space with 3/2 dealiasing and explicit Euler finite differences in time. The stochastic coefficients and translation amount are integrated in time with respectively a 4th-order Runge-Kutta and a 2-step Adams-Bashforth scheme, both using a particle method. Note that setting the stochastic translation amount to zero in the SDO equations readily yields the standard DO scheme. For both SDO and DO, we use $L_1 = L_2 = 2\pi$, $64 \times 64$ Fourier modes, $\Delta t = 0.001$ and 1000 Monte-Carlo particles.

\subsection{Example on a stochastically advected vortex}
\label{sec:ExampleVortex}

We now use the SDO equations to compute the transient flow of a randomly advected stochastic Lamb-Oseen vortex, described by the following initial velocity field
\begin{equation}
\mathbf{u}_0(x_1,x_2;\omega) = \frac{\Gamma}{2\pi r} \left[1-\exp \left( -\frac{r^2}{r_c^2(\omega)} \right) \right] \left( -\sin \theta, \cos \theta \right)^\mathsf{T} + \left(\cos \psi(\omega), \sin \psi(\omega) \right)^\mathsf{T},
\label{eq:LambOseen}
\end{equation}
where $r = \sqrt{(x_1-L_1/2)^2+(x_2-L_2/2)^2}$ and $\theta = \arctan x_2/x_1$. In the above initial condition, a Lamb-Oseen vortex centered in the domain with random radial length scale $r_c(\omega) \sim \mathcal{N}(0.2,10^{-4})$ is superimposed to a uniform flow of unit magnitude and random direction $\psi(\omega) \sim \mathcal{U}(0,\pi/2)$. We take $\Gamma = 10$ and set the Reynolds number to $Re = 40$. Each realization $\omega$ defined by the above initial condition therefore undergoes viscous diffusion while being simultaneously advected in a given direction due to the uniform flow. Thus, after a finite time, the realizations will have moved away from their initial position to various locations in the domain. 

\begin{figure}
\centering
\includegraphics[width=\textwidth]{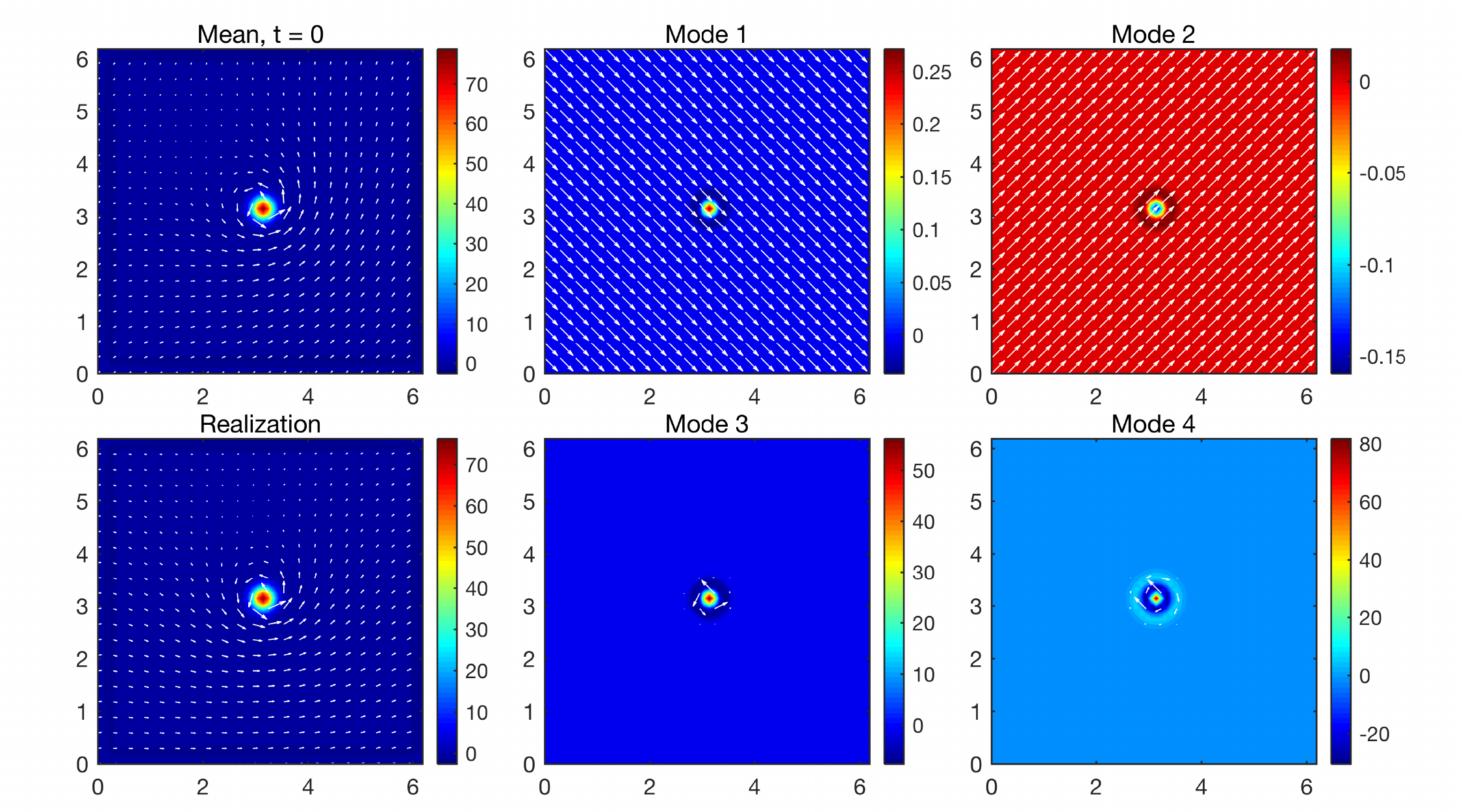}
\caption{Initial conditions for the mean and first four modes of the SDO and DO simulations, pictured in terms of their vorticity (background color) and velocity (arrows) fields. Due to the fact that the realizations defined by \eqref{eq:LambOseen} are centered in the domain, the initial conditions are identical between the SDO and DO computations. A single initial realization is also shown.}
\label{fig:SDOIC}
\end{figure}

The SDO simulation is initialized according to the procedure described in Section \ref{sec:InitializationSDONS}. For comparison purposes, we also perform a regular DO simulation of the same problem and we use 6 modes in both computations. Since \eqref{eq:LambOseen} defines the initial realizations as centered in the domain, they are initially identical to their symmetry-reduced counterparts which are also required to be centered in the domain. It therefore follows that the initial conditions for the SDO and DO mean and modes are the same, and they are displayed in Figure \ref{fig:SDOIC} in terms of their vorticity (background color) and velocity (arrows) fields.
Notice that the first two modes mainly represent the variability due to the uniform advection flow of unit magnitude and random direction, whereas the subsequent modes account for the variability due to the random radial length scale of the Lamb-Oseen vortex. 

\begin{figure}
\centering
\includegraphics[width=\textwidth]{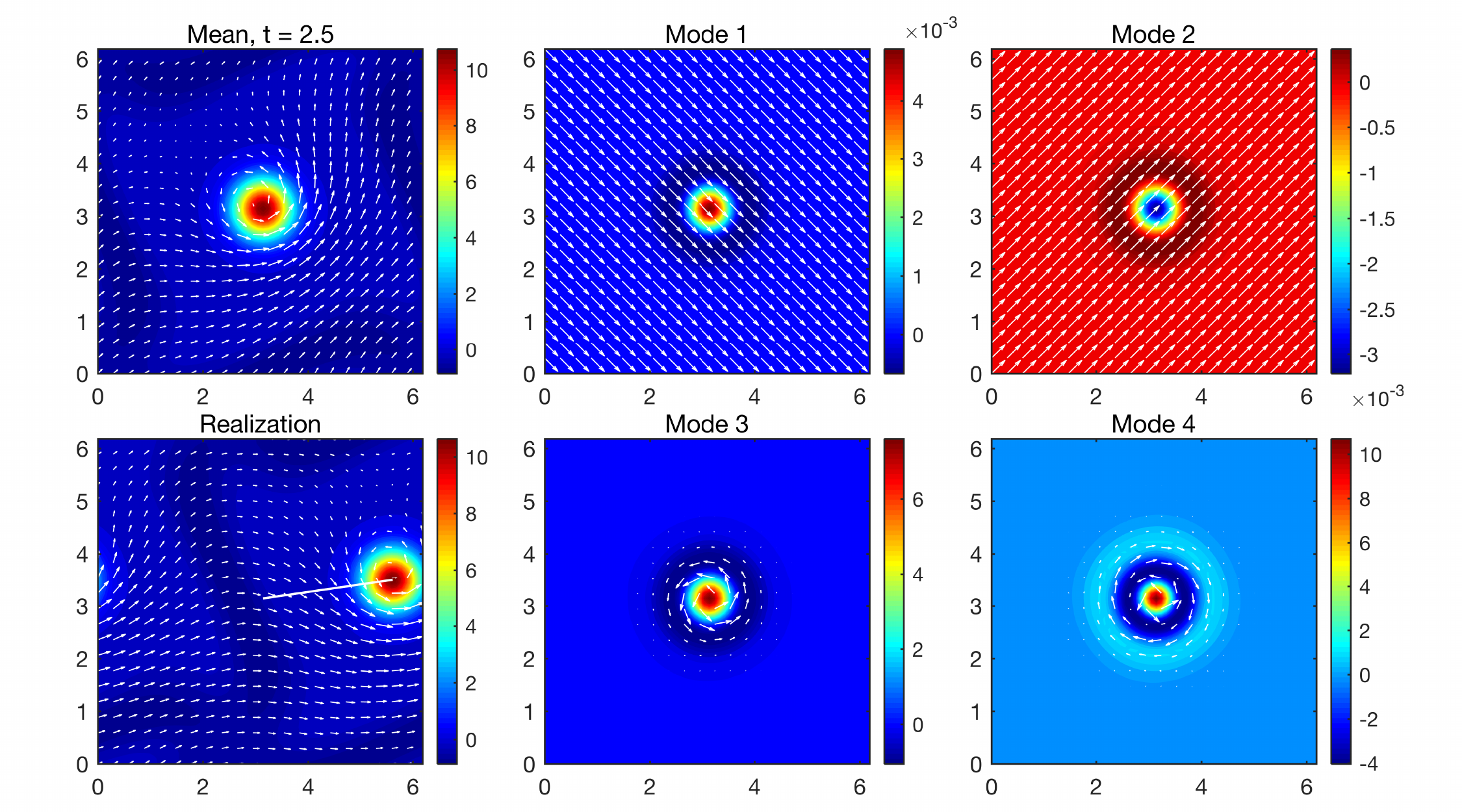}
\caption{Symmetry-reduced mean $\bar{\mathbf{u}}$, modes $\hat{\mathbf{u}}_i$ and a realization of the SDO solution at final time $t = 2.5$, pictured in terms of their vorticity (background color) and velocity (arrows) fields. The realization is recovered from its symmetry-reduced state as $\mathbf{u} = g(\mathbf{c}) \hat{\mathbf{u}}$, and the white line shows its trajectory obtained from the time history of the translation amount $\mathbf{c}$.}
\label{fig:SDOQuart}
\end{figure}

Figure \ref{fig:SDOQuart} shows the symmetry-reduced mean $\bar{\mathbf{u}}(\mathbf{x},t)$ and symmetry-reduced modes $\hat{\mathbf{u}}_i(\mathbf{x},t)$ of the SDO solution at time $t = 2.5$, together with one full state space realization reconstructed from the SDO quantities through the group transformation \eqref{eq:ReconstructionNS}.
Compared with the initial condition shown in Figure \ref{fig:SDOIC}, we observe that the symmetry-reduced mean and modes have remained at their initial location and have merely diffused out due to viscosity. On the other hand, the realization has not only diffused out but has moved away from the center of the domain along the path shown by the white line, thanks to the stochastic shift amount $\mathbf{c}(t;\omega)$ which has tracked the advection due to the stochastic uniform flow. 

\begin{figure}
\centering
\includegraphics[width=0.33\textwidth]{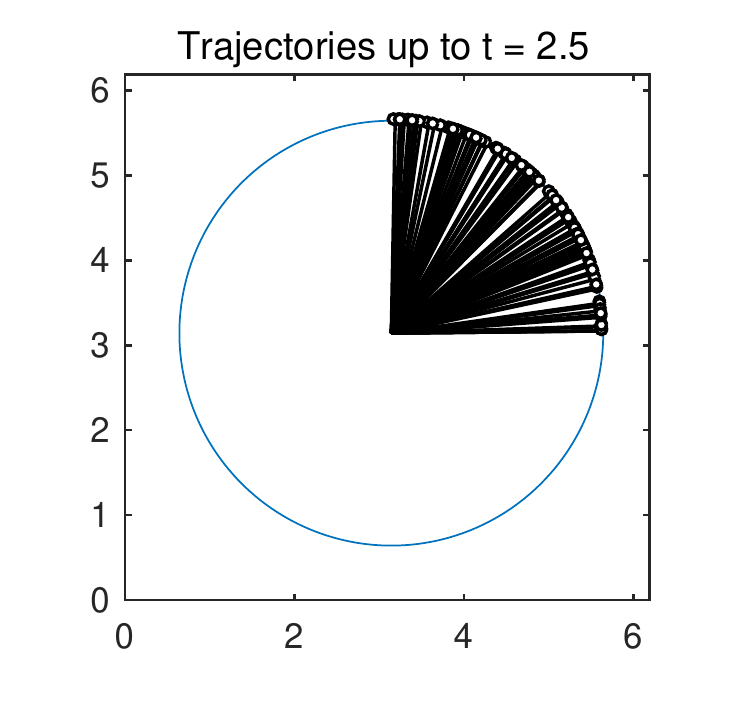}
\includegraphics[width=0.5\textwidth]{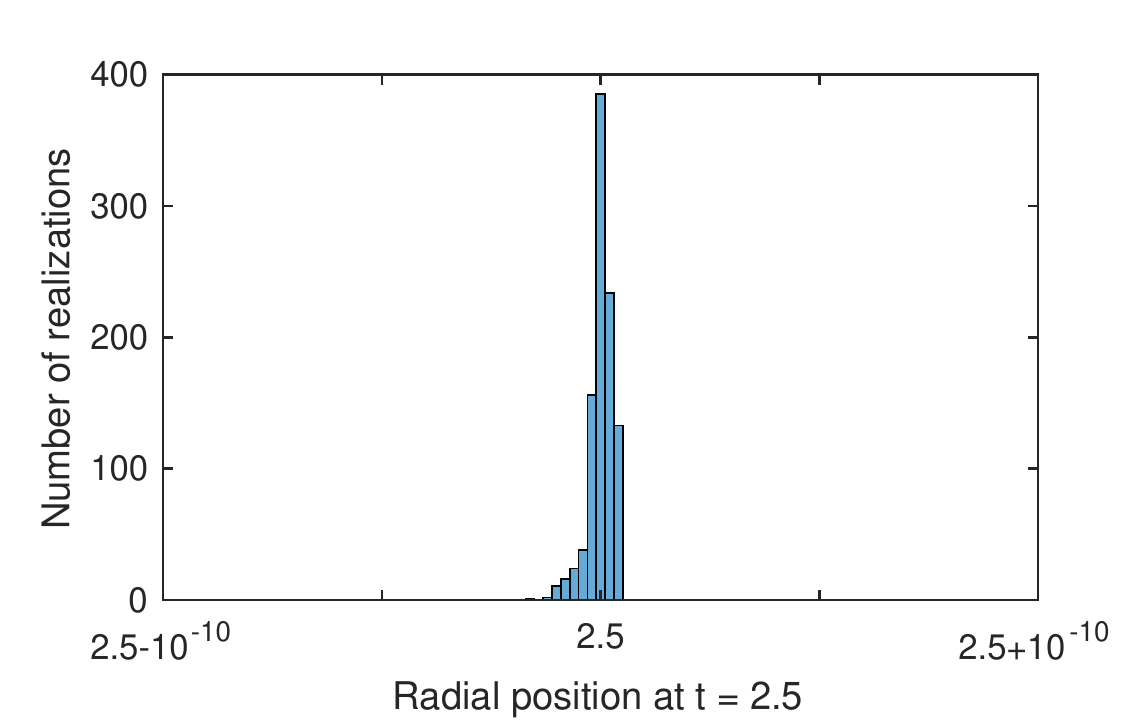}
\caption{Left: Trajectories of 100 sample realizations of the SDO solution between initial and final time, obtained from the time history of the stochastic translation amount $\mathbf{c}(t;\omega)$. Right: Distribution of the radial distance $||\mathbf{c}(t;\omega)||$ of the realizations from their initial position at final time $t = 2.5$.}
\label{fig:TrajectoriesSDOQuart}
\end{figure}

In the left-hand side of Figure \ref{fig:TrajectoriesSDOQuart}, we further display the individual trajectories of 100 realizations, obtained from the time history of the shift amount.
The realizations move away from their initial position at unit velocity and in a random direction uniformly distributed between 0 and $\pi/2$, eventually landing at final time on a circle of radius 2.5. These trajectories are expected given the initial condition \eqref{eq:LambOseen}, but nonetheless remarkable since the SDO equations do not have \textit{a priori} knowledge of the paths followed by the realizations. The exact radial distance $||\mathbf{c}(t;\omega)||$ of the realizations to their initial position at final time is shown in the right-hand side of Figure \ref{fig:TrajectoriesSDOQuart} and has negligible standard deviation on the order of $10^{-11}$, which demonstrates the numerical accuracy of the stochastic shift amount. 

\begin{figure}
\centering
\includegraphics[width=\textwidth]{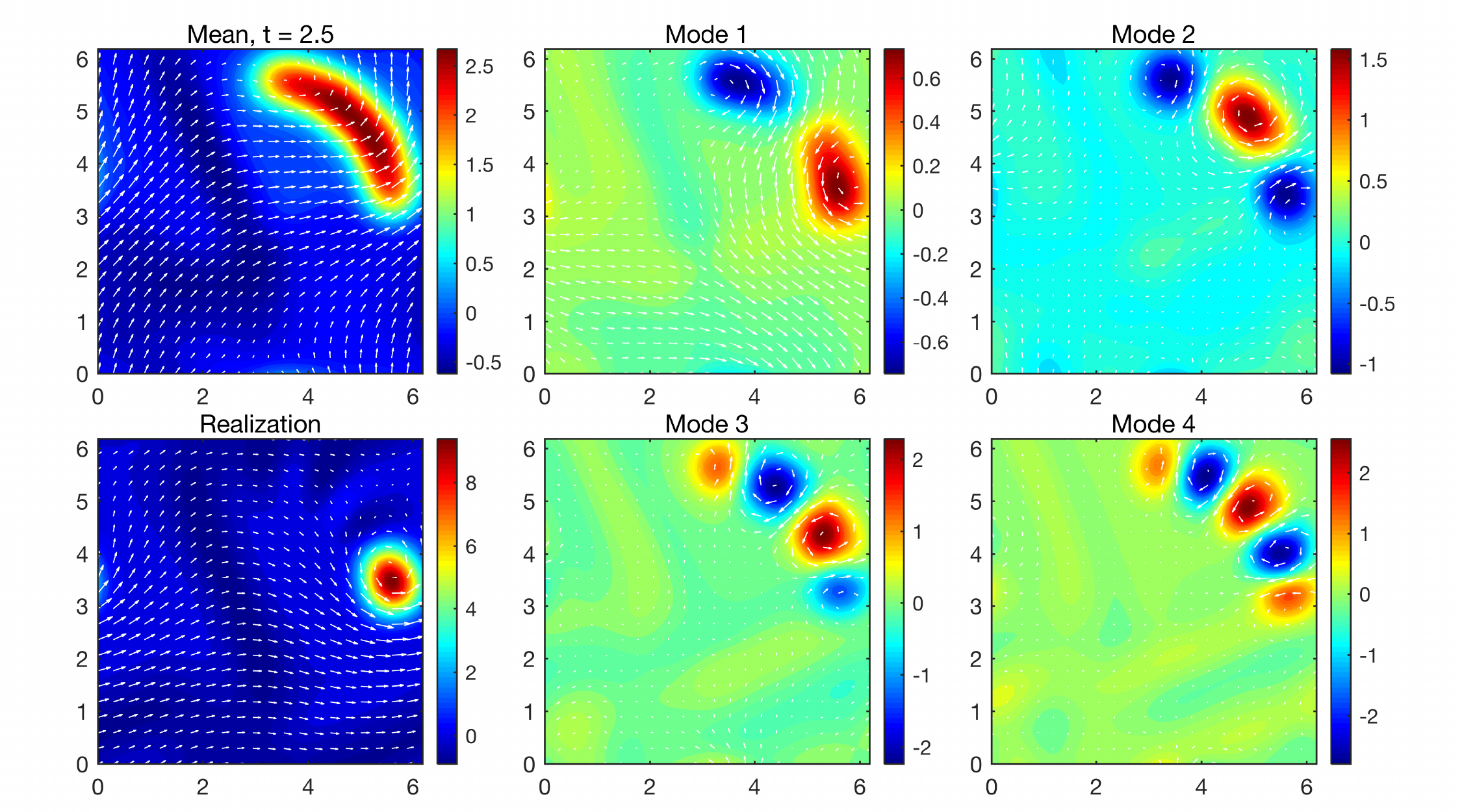}
\caption{Mean $\bar{\mathbf{u}}$, modes $\mathbf{u}_i$ and a realization of the DO solution at final time $t = 2.5$, pictured in terms of their vorticity (background color) and velocity (arrows) fields. The mean and modes directly correspond to the low-dimensional projection of the full state space solution.}
\label{fig:DOQuart}
\end{figure}

For comparison purposes, Figure \ref{fig:DOQuart} shows the mean $\bar{\mathbf{u}}(\mathbf{x},t)$ and modes $\mathbf{u}_i(\mathbf{x},t)$ of the DO solution at time $t = 2.5$, together with one realization.
Since the standard DO scheme directly performs order reduction in the full state space, the DO mean and modes need to account for the spatial dispersion of the realizations. As a consequence, the modes in Figure \ref{fig:DOQuart} do not directly reflect the shape of each individual vortex and a limited number of them is unable to accurately reproduce each realization. On the contrary, the SDO modes are physically more relevant since they directly reveal the underlying non-trivial dynamics of the solution.

\begin{figure}
\centering
\includegraphics[width=0.49\textwidth]{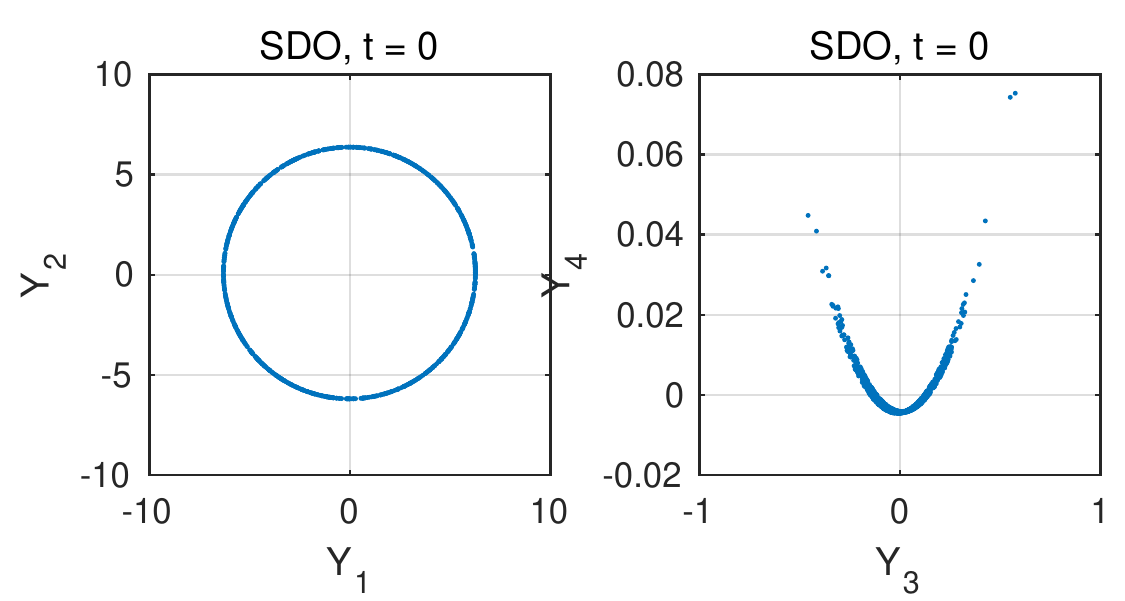}
\includegraphics[width=0.49\textwidth]{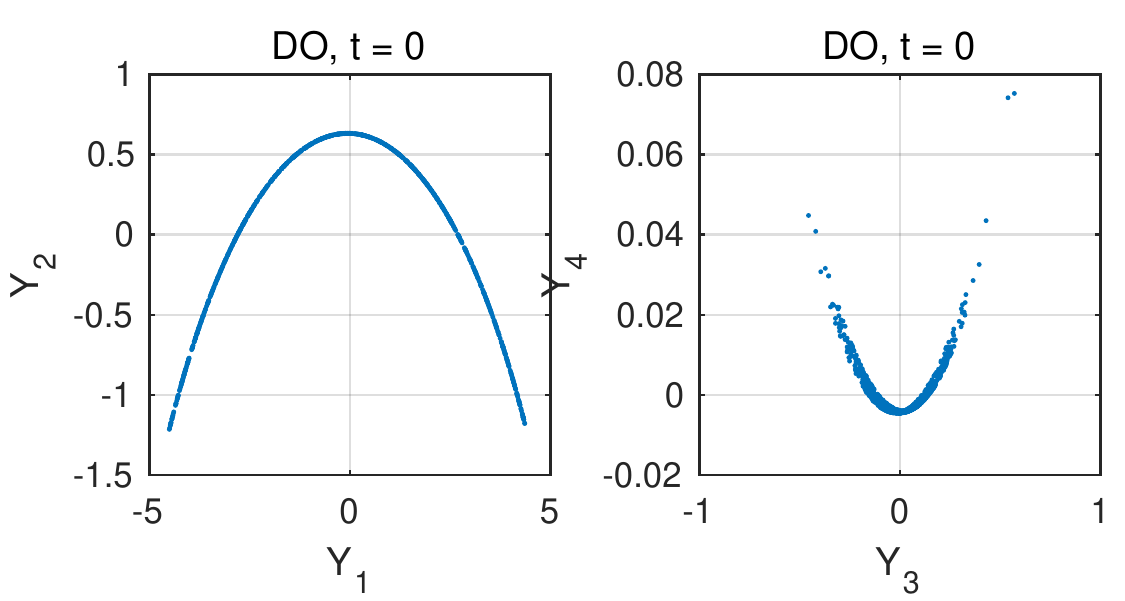}
\includegraphics[width=0.49\textwidth]{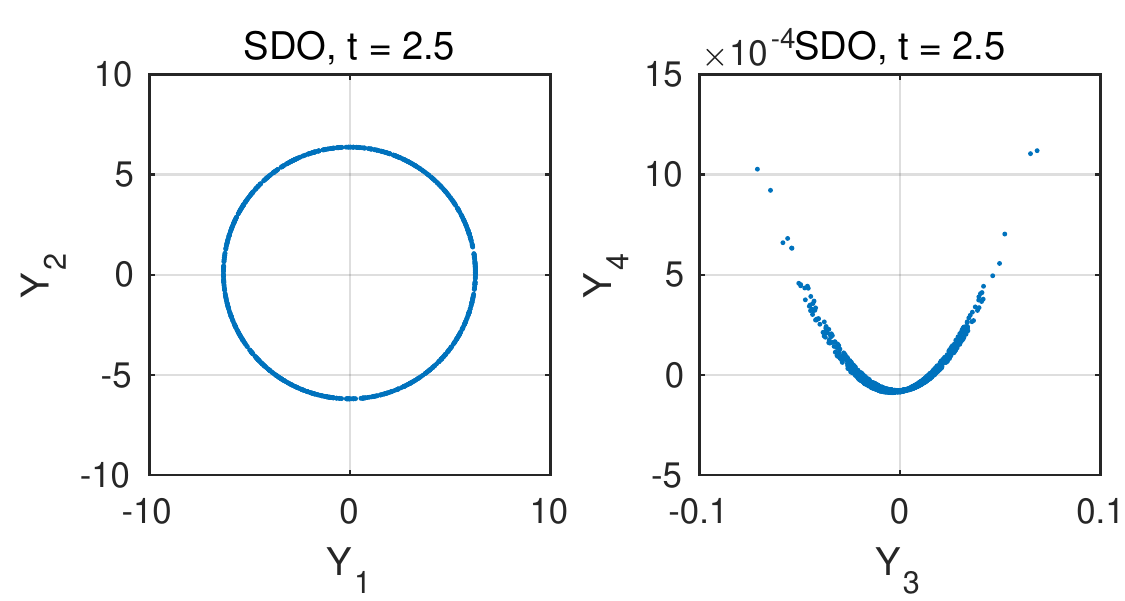}
\includegraphics[width=0.49\textwidth]{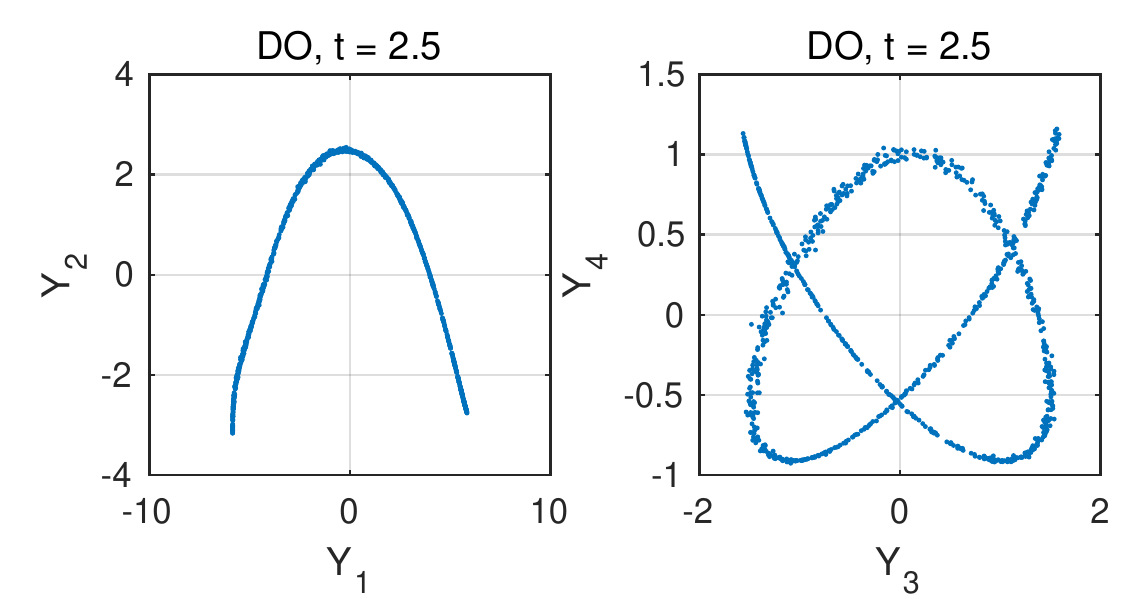}
\caption{Statistics of the first four SDO (left) and DO (right) stochastic coefficients at initial and final times, displayed in terms of the two-dimensional marginals.}
\label{fig:JointDistributionQuart}
\end{figure}

Figure \ref{fig:JointDistributionQuart} shows the statistics of the first four SDO (left) and DO (right) stochastic coefficients at initial and final times, displayed in terms of two-dimensional marginals.
The statistical structure of the SDO solution remains identical over time except for some decline in the variance of $Y_3$ and $Y_4$ due to viscous diffusion. On the other hand, the statistical structure of the DO solution completely changes over time since the modes have to account for the spatial translation of each realization, thereby obscuring the fact that each individual vortex is merely diffusing besides the advection.

\begin{figure}
\centering
\includegraphics[width=0.49\textwidth]{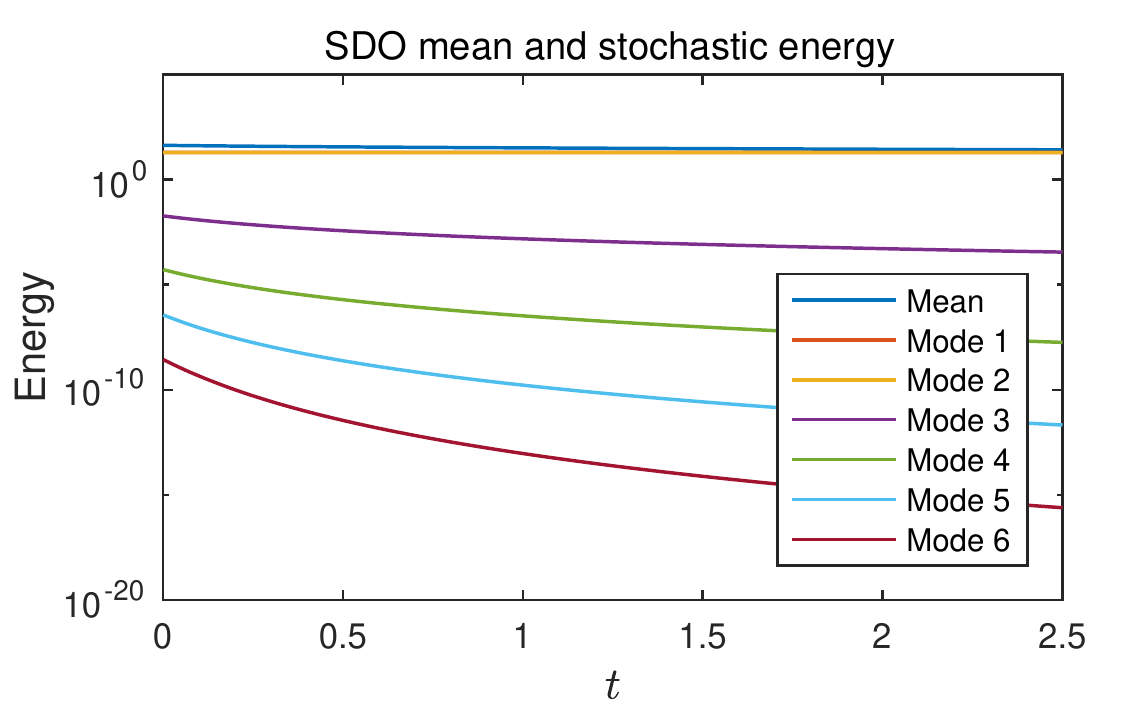}
\includegraphics[width=0.49\textwidth]{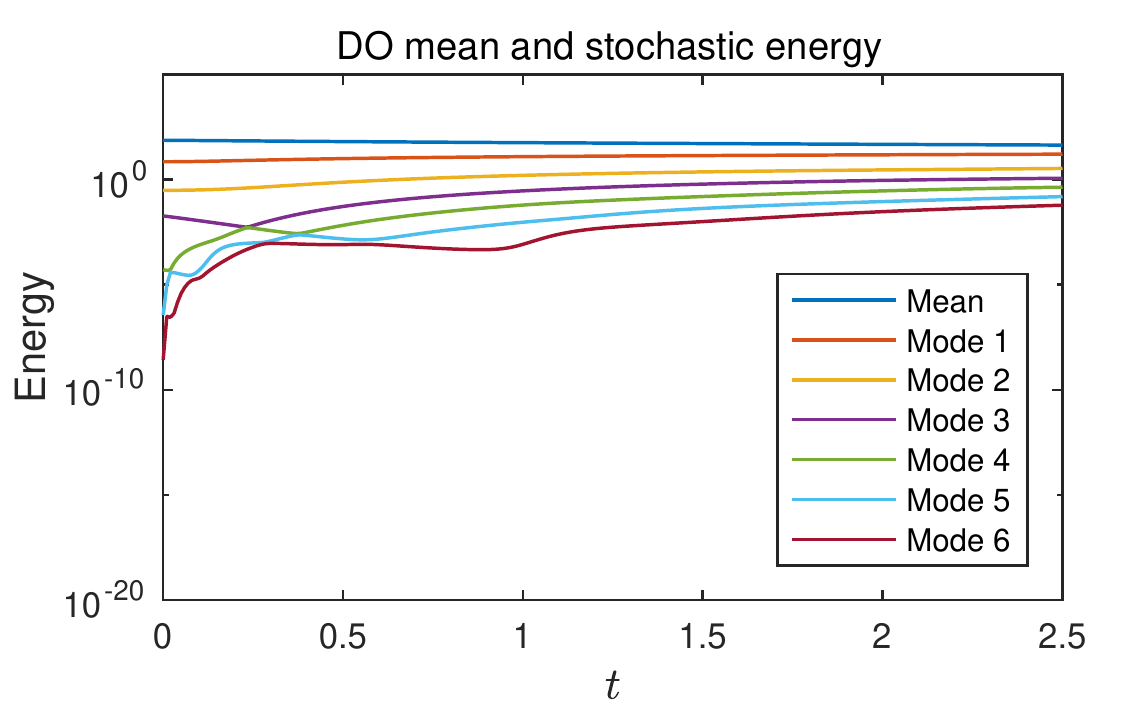}
\caption{Time evolution of the energy in the mean and variance of the stochastic coefficients of the SDO (left) and DO (right) solutions.}
\label{fig:VarianceQuart}
\end{figure}

\begin{figure}
\centering
\includegraphics[width=0.49\textwidth]{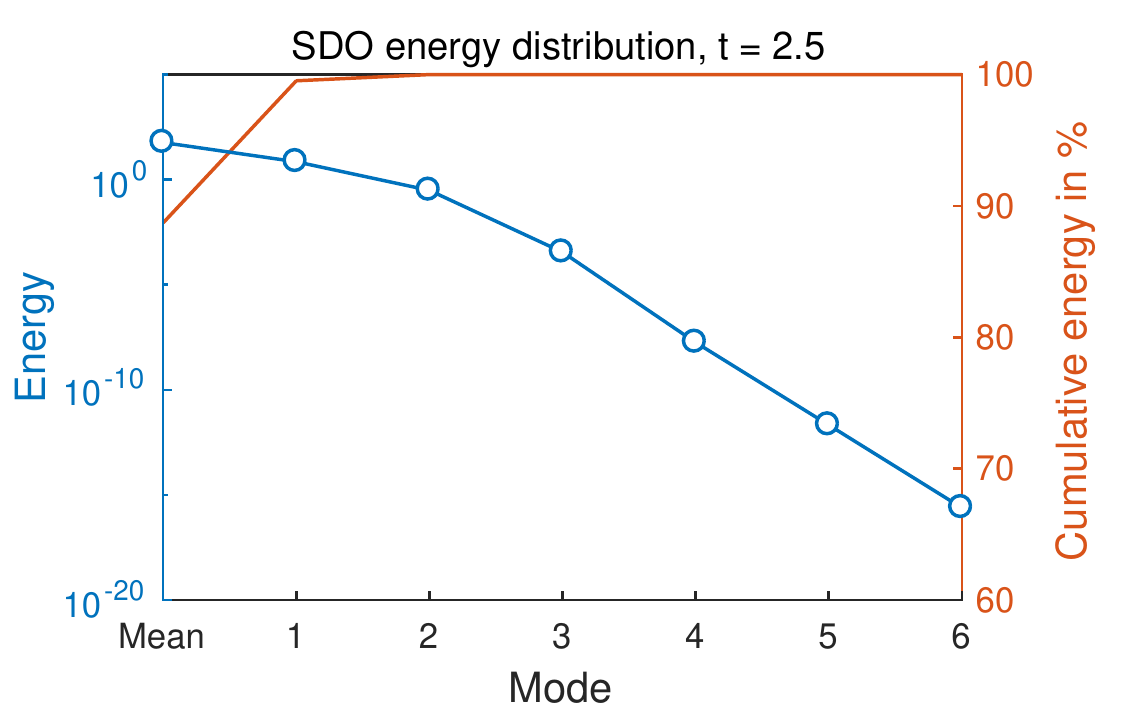}
\includegraphics[width=0.49\textwidth]{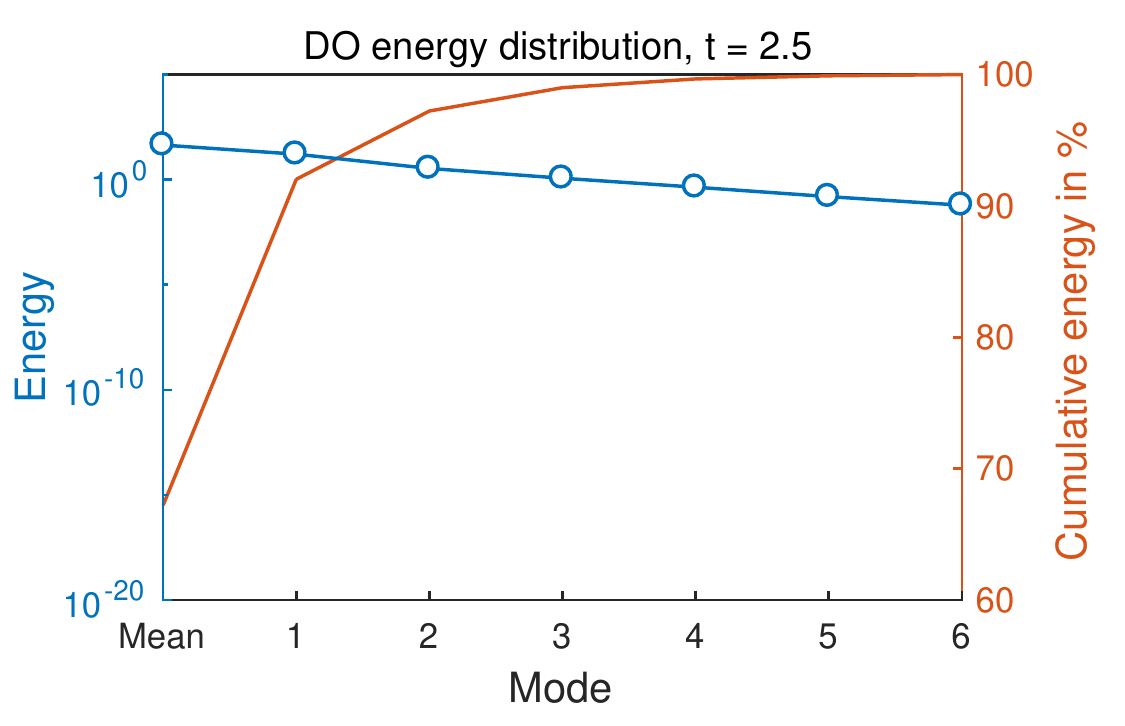}
\caption{Individual and cumulative distribution of the energy in the mean and variance of the stochastic coefficients of the SDO (left) and DO (right) solutions at final time $t = 2.5$.}
\label{fig:EnergyDistributionQuart}
\end{figure}

This difference in the amount of stochastic variability accounted for by the modes is clearly apparent in Figure \ref{fig:VarianceQuart}, which displays the time evolution of the energy in the mean $\langle \bar{\mathbf{u}},\bar{\mathbf{u}} \rangle$ and the variance of the stochastic coefficients $E[Y_i^2]$ for the SDO (left) and DO (right) solutions. 
The variance of the SDO mean and modes decreases over time since the symmetry-reduced state simply undergoes viscous diffusion. On the other hand, that of their DO counterparts grows over time as the realizations become more and more dispersed. Therefore, the SDO scheme achieves much better accuracy than DO with the same number of modes, as can also be seen from the individual and cumulative distributions of energy in the mean and the modes at final time $t = 2.5$ shown in Figure \ref{fig:EnergyDistributionQuart}.

These simulations are shown as videos in the Supplementary Materials, together with a second case where the direction of the uniform advection flow in the initial condition is uniformly distributed between 0 and $2\pi$, instead of between 0 and $\pi/2$.

\section{Conclusions}
\label{sec:Conclusions}

In this work, we have introduced a novel methodology for efficient order reduction of stochastic dynamical systems with continuous symmetries. This methodology is composed of two steps. In a first step, one performs symmetry reduction of the original dynamical system using the method of slices. In this way, the dynamics of the original system is decoupled into shape deformations, captured by a symmetry-reduced stochastic state fixed in the physical domain, and motion along the symmetry directions of the system, tracked by a set of scalar phase parameters. The second step consists in order reduction of the symmetry-reduced stochastic state, using any standard order reduction method of choice. Since the symmetry-reduced state is fixed in the physical domain, this procedure results in very efficient mixed symmetry-dimensionality reduction schemes. In particular, using the Dynamically Orthogonal (DO) equations to perform the second step, we have obtained a new Symmetry-reduced Dynamically Orthogonal (SDO) scheme that shows much better performance than DO on stochastic solutions of the 1D Korteweg-de Vries and 2D Navier-Stokes equations. 

Even though both examples we have presented consider translation invariance and utilize periodic boundary conditions, it is important to recognize that continuous symmetries can arise in bounded domains as well. A prominent example is axisymmetric systems, for instance the flow past an axisymmetric body or in a Taylor-Couette cell. Using axisymmetric coordinates, the rotational symmetry translates to translational invariance in the azimuthal direction, allowing for the direct application of our technique. Furthermore, while we have only considered stochasticity through the initial conditions, we expect our framework to be equally applicable to stochastic operators provided that the stochastic terms (either coefficients or forcing) are statistically homogeneous in the symmetry direction, ensuring that the equivariance condition \eqref{eq:EquivarianceCondition} is satisfied in a statistical sense.

In general, we expect this methodology to work well whenever the stochastic solution is composed of coherent structures or sharp gradients like in shock waves. Possible future work could include the extension of this framework to dynamical systems with self-similar solutions, which can also be symmetry-reduced with the method of slices by rescaling time as well \cite{aronson2001,rowley2003,siettos2003}. Finally, one could also envision complementing our method with data assimilation techniques such as Bayesian inference, to which group transformation ideas have already been applied \cite{ravela2007}.

\section*{Acknowledgments}
The authors would like to thank Dr.~Edouard Boujo for a careful reading of the manuscript, the two anonymous reviewers for their constructive comments, as well as Dr.~Sai Ravela, Dr.~Mohammad Farazmand and Florian Feppon for helpful discussions related to the subject. This work has been supported through the ONR grant N00014-15-1-2381, the AFOSR grants FA9550-16-1-0231 and W911NF-17-1-0306, and the ARO grant 66710-EG-YIP.

\appendix

\section{DO order reduction preserves symmetry reduction}
\label{app:OrderReductionPreservesSymmetryReduction}

In this appendix, we show that after DO order reduction of the symmetry-reduced equations, the dynamics of the reduced-order solution remain confined to the symmetry-reduced state space. First, assume that the phase parameters $\boldsymbol{\phi}$ satisfy the stochastic equation \eqref{eq:DOPhaseParameters}, which implies
\begin{equation}
\langle \mathbf{F}(\hat{\mathbf{u}}) - \dot{\phi}_a \mathbf{t}_a(\hat{\mathbf{u}}), \mathbf{t}_b' \rangle = 0,
\end{equation}
for all symmetry directions $b = 1,...,N$ and all reduced-order realizations \eqref{eq:DORepresentation}. We also assume that at the current time instant, the mean and the modes satisfy the slice condition \eqref{eq:SliceCondition2}, that is $\langle \bar{\mathbf{u}}, \mathbf{t}_a' \rangle = \langle \hat{\mathbf{u}}_i, \mathbf{t}_a' \rangle = 0$. Then, taking the inner products of the evolution equation \eqref{eq:DOMean} for the mean with the template tangents, we find that
\begin{equation}
\langle \frac{\partial \bar{\mathbf{u}}}{\partial t}, \mathbf{t}_b' \rangle = \langle E[\mathbf{F}(\hat{\mathbf{u}}) - \dot{\phi}_a \mathbf{t}_a(\hat{\mathbf{u}})], \mathbf{t}_b' \rangle = E[\langle \mathbf{F}(\hat{\mathbf{u}}) - \dot{\phi}_a \mathbf{t}_a(\hat{\mathbf{u}}), \mathbf{t}_b' \rangle] = 0,
\label{eq:DOMeanTemplateTangents}
\end{equation}
for $b = 1,...,N$. Likewise, taking the inner products of the evolution equation \eqref{eq:DOModes} for the modes with the template tangents, we have
\begin{equation}
\langle \frac{\partial \hat{\mathbf{u}}_i}{\partial t}, \mathbf{t}_b' \rangle = \langle \hat{\mathbf{H}}_i, \mathbf{t}_b' \rangle - \langle \hat{\mathbf{H}}_i, \hat{\mathbf{u}}_j \rangle \langle \hat{\mathbf{u}}_j, \mathbf{t}_b' \rangle = 0,
\label{eq:DOModesTemplateTangents}
\end{equation}
where we have used $\langle \hat{\mathbf{u}}_j, \mathbf{t}_b' \rangle = 0$ and the fact that
\begin{equation}
\langle \hat{\mathbf{H}}_i, \mathbf{t}_b' \rangle = \langle E[Y_k(\mathbf{F}(\hat{\mathbf{u}}) - \dot{\phi}_a \mathbf{t}_a(\hat{\mathbf{u}}))] \, C_{ik}^{-1}, \mathbf{t}_b' \rangle = E[Y_k C_{ik}^{-1} \langle \mathbf{F}(\hat{\mathbf{u}}) - \dot{\phi}_a \mathbf{t}_a(\hat{\mathbf{u}}), \mathbf{t}_b' \rangle ] = 0.
\end{equation}
As in \eqref{eq:DynamicSliceCondition}, equations \eqref{eq:DOMeanTemplateTangents} and \eqref{eq:DOModesTemplateTangents} show that the mean and modes will satisfy the slice condition \eqref{eq:SliceCondition2} throughout time integration, so that the reduced-order solution \eqref{eq:DORepresentation} remains confined to the symmetry-reduced state space. Numerical round-off errors remain negligible in our simulations.

\section{First Fourier mode slice for the KdV equation}
\label{app:FirstFourierModeSliceKdVEquation}

Here, we justify that the template function \eqref{eq:TemplateKdV} corresponds to the first Fourier mode slice defined in Section \ref{sec:ChoiceTemplateKdV}, that is the slice condition $\langle \hat{u}, t(\hat{u}') \rangle = 0$ is satisfied when the phase angle of the first Fourier mode of $\hat{u}$ is equal to $\pi$. First, note that the inner product between two generic states $u(x)$ and $v(x)$ can be expressed in terms of their Fourier coefficients $\tilde{u}(k) = a(k) + i b(k)$ and $\tilde{v}(k) = c(k) + i d(k)$ as
\begin{equation}
\langle u, v \rangle = L \left[ \tilde{u}(0) \tilde{v}(0) + 2 \sum_{k = 1}^\infty \left( a(k) c(k) + b(k) d(k) \right) \right],
\label{eq:InnerProductFourier}
\end{equation}
where we have used that $\tilde{u}(-k) = \tilde{u}^*(k)$ and $\tilde{v}(-k) = \tilde{v}^*(k)$ since $u$ and $v$ are real. If we now denote $\tilde{\hat{u}}(k) = a(k) + i b(k)$ and $\tilde{\hat{u}}'(k) = c(k) + i d(k)$ the Fourier coefficients of the symmetry-reduced state $\hat{u}$ and template $\hat{u}'$, respectively, we can use \eqref{eq:InnerProductFourier} to express the slice condition as
\begin{equation}
\langle \hat{u}, t(\hat{u}') \rangle = \langle \hat{u}, -\partial_x \hat{u}' \rangle = 4 \pi \sum_{k = 1}^\infty \left( k a(k) d(k) - k b(k) c(k) \right) = 0.
\label{eq:SliceConditionKdV}
\end{equation}
The first Fourier mode slice is defined by $\arg \tilde{\hat{u}}(1) = \pi$, thus $b(1) = 0$, $a(1) < 0$ and all other coefficients $a(k)$, $b(k)$ are possibly non-zero. Except for the sign of $a(1)$, these conditions are equivalent to the above slice condition when $c(1) \neq 0$ and all other coefficients $c(k)$, $d(k)$ are zero. Hence a valid choice of template is
\begin{equation}
\hat{u}' = \cos \frac{2 \pi x}{L},
\end{equation}
which corresponds to $c(1) = 1/2$. With this template, the slice condition \eqref{eq:SliceConditionKdV} reduces to $\langle \hat{u}, t(\hat{u}') \rangle = -2 \pi b(1) = 0$, which is equivalent to requiring that $\arg \tilde{\hat{u}}(1) = 0$ or $\pi$. This ambiguity between zero and $\pi$ is not a problem in the actual SDO computations since the symmetry-reduced state $\hat{u}$ is initialized with $\arg \tilde{\hat{u}}(1,t_0;\omega) = \pi$, and continuity of the phase angle during time integration ensures that $\arg \tilde{\hat{u}}(1,t;\omega)$ does not jump to zero at later times. With the phase of its first Fourier mode fixed, the symmetry-reduced stochastic state $\hat{u}(x,t;\omega)$ is effectively pinned at a fixed location in space and this is accomplished with a generic template that is not problem-dependent.

\section{First Fourier mode slice for the Navier-Stokes equation}
\label{app:FirstFourierModeSliceNSEquation}

Here, we justify that template \eqref{eq:TemplateNS} corresponds to the first Fourier mode slice defined in Section \ref{sec:ChoiceTemplateNS} for a two-dimensional velocity field, that is, both slice conditions $\langle \hat{\mathbf{u}},\mathbf{t}_1(\hat{\mathbf{u}}') \rangle = 0$ and $\langle \hat{\mathbf{u}},\mathbf{t}_2(\hat{\mathbf{u}}') \rangle = 0$ are satisfied when the phase angles of the first Fourier modes of $\hat{\omega}$ in the $x$ and $y$ directions are equal to $\pi$. First, note that the inner product between two generic states $\mathbf{u}(\mathbf{x}) = (u_1(x_1,x_2),u_2(x_1,x_2))^\mathsf{T}$ and $\mathbf{v}(\mathbf{x}) = (v_1(x_1,x_2),v_2(x_1,x_2))^\mathsf{T}$ can be expressed in terms of their Fourier coefficients $\tilde{\mathbf{u}}(\mathbf{k}) = (\tilde{u}_1(k_1,k_2),\tilde{u}_2(k_1,k_2))^\mathsf{T}$ and $\tilde{\mathbf{v}}(\mathbf{k}) = (\tilde{v}_1(k_1,k_2),\tilde{v}_2(k_1,k_2))^\mathsf{T}$ as
\begin{align}
\langle \mathbf{u}, \mathbf{v} \rangle &= L_1 L_2 \Bigg[ \tilde{u}_1(0,0) \tilde{v}_1(0,0) + \tilde{u}_2(0,0) \tilde{v}_2(0,0) \nonumber \\
&\quad + 2 \sum_{k_1,k_2 = 1}^\infty \left( a_1(k_1,k_2) c_1(k_1,k_2) + b_1(k_1,k_2) d_1(k_1,k_2) \right) \nonumber \\
&\quad + 2 \sum_{k_1,k_2 = 1}^\infty \left( a_2(k_1,k_2) c_2(k_1,k_2) + b_2(k_1,k_2) d_2(k_1,k_2) \right) \Bigg], 
\label{eq:InnerProductFourier2D} 
\end{align}
where $\tilde{u}_j(k_1,k_2) = a_j(k_1,k_2) + i b_j(k_1,k_2)$ and $\tilde{v}_j(k_1,k_2) = c_j(k_1,k_2) + i d_j(k_1,k_2)$, $j = 1,2$, and we have used the realness of $u_j$ and $v_j$. If we now denote $\tilde{\hat{\mathbf{u}}}(\mathbf{k}) = (\tilde{\hat{u}}_1(k_1,k_2),\tilde{\hat{u}}_2(k_1,k_2))^\mathsf{T}$ and $\tilde{\hat{\mathbf{u}}}'(\mathbf{k}) = (\tilde{\hat{u}}'_1(k_1,k_2),\tilde{\hat{u}}'_2(k_1,k_2))^\mathsf{T}$ the Fourier coefficients of the symmetry-reduced state $\hat{\mathbf{u}}$ and template $\hat{\mathbf{u}}'$, respectively, we can use \eqref{eq:InnerProductFourier2D} to express the slice conditions $\langle \hat{\mathbf{u}},\mathbf{t}_a(\hat{\mathbf{u}}') \rangle = 0$, $a = 1,2$ as
\begin{align}
\langle \hat{\mathbf{u}}, -\partial_{x_a} \hat{\mathbf{u}}' \rangle &= 4 \pi \frac{L_1 L_2}{L_a} \Bigg[ \sum_{k_1,k_2 = 1}^\infty \left( k_a a_1(k_1,k_2) d_1(k_1,k_2) - k_a b_1(k_1,k_2) c_1(k_1,k_2) \right) \nonumber \\
&\quad + \sum_{k_1,k_2 = 1}^\infty \left( k_a a_2(k_1,k_2) d_2(k_1,k_2) - k_a b_2(k_1,k_2) c_2(k_1,k_2) \right) \Bigg] \nonumber \\
&= 0,
\label{eq:SliceConditionsNS}
\end{align}
where $\tilde{\hat{u}}_j(k_1,k_2) = a_j(k_1,k_2) + i b_j(k_1,k_2)$ and $\tilde{\hat{u}}'_j(k_1,k_2) = c_j(k_1,k_2) + i d_j(k_1,k_2)$, $j = 1,2$. From Section \ref{sec:ChoiceTemplateNS}, one can show that the first Fourier mode slice corresponds to $\arg \tilde{\hat{u}}_1(0,1) = -\pi/2$ and $\arg \tilde{\hat{u}}_2(1,0) = \pi/2$, thus $a_1(0,1) = a_2(1,0) = 0$, $b_1(0,1) < 0$, $b_2(1,0) > 0$ and all other coefficients $a_j(k_1,k_2)$, $b_j(k_1,k_2)$, $j = 1,2$ are possibly non-zero. Except for the signs of $b_1(0,1)$ and $b_2(1,0)$, these conditions are equivalent to the slice conditions \eqref{eq:SliceConditionsNS} when $d_1(0,1) \neq 0$, $d_2(1,0) \neq 0$ and all other coefficients $c_j(k_1,k_2)$, $d_j(k_1,k_2)$, $j = 1,2$ are zero. Hence a valid choice of template is
\begin{equation}
\hat{\mathbf{u}}' = \left( \sin \frac{2 \pi x_2}{L_2}, \sin \frac{2 \pi x_1}{L_1} \right)^\mathsf{T},
\end{equation}
which corresponds to $d_1(0,1) = d_2(1,0) = -1/2$. With this template, the slice conditions \eqref{eq:SliceConditionsNS} become $\langle \hat{\mathbf{u}}, \mathbf{t}_1(\hat{\mathbf{u}}') \rangle = -2 \pi L_2 a_2(1,0) = 0$ and $\langle \hat{\mathbf{u}}, \mathbf{t}_2(\hat{\mathbf{u}}') \rangle = -2 \pi L_1 a_1(0,1) = 0$, which means requiring that $\arg \tilde{\hat{u}}_2(1,0) = \arg \tilde{\hat{u}}_1(0,1) = \pm \pi/2$. Same as for the one-dimensional case, this ambiguity between $\pm \pi/2$ is not an issue in the actual SDO computations since the symmetry-reduced state $\hat{\mathbf{u}}$ is initialized according to the first Fourier mode slice with $\arg \tilde{\hat{u}}_1(0,1,t_0;\omega) = -\pi/2$ and $\arg \tilde{\hat{u}}_2(1,0,t_0;\omega) = \pi/2$. Continuity of the phase angles during time integration then ensures that $\arg \tilde{\hat{u}}_1(0,1,t;\omega)$ and $\arg \tilde{\hat{u}}_2(1,0,t;\omega)$ do not jump to the other root at later times.

\bibliography{References}

\end{document}